\documentclass[aps,prb,reprint,superscriptaddress]{revtex4-1}

\usepackage{bm}
\usepackage{graphicx}
\usepackage{comment}
\begin{document}
\newcommand{\mub}{${\,}\mu_\textrm{\scalebox{0.5}{B}}\,$}
\newcommand{\tm}{$T_\textrm{\scalebox{0.6}{m}}\,$}

\title{Peculiar Magnetism of UAu$_{2}$Si$_{2}$}

\author{Chihiro Tabata}
\altaffiliation[Present address: ]{Condensed Matter Research Center and Photon Factory,
Institute of Materials Structure Science, High Energy Accelerator Research Organization, Tsukuba, Japan}
\email[E-mail: ]{ctabata@post.kek.jp}
\affiliation{Graduate School of Science, Hokkaido University, Sapporo 060-0810, Japan}
\author{Naoyuki Miura}
\affiliation{Graduate School of Science, Hokkaido University, Sapporo 060-0810, Japan}
\author{Kl{\'a}ra~\surname{Uhl{\'i}{\v{r}}ov{\'a}}}
\author{Michal~\surname{Vali{\v{s}}ka}}
\affiliation{Faculty of Mathematics and Physics, Charles University, Ke Karlovu 5, 121 16 Prague 2, Czech Republic}
\author{Hiraku Saito}
\author{Hiroyuki Hidaka}
\author{Tatsuya Yanagisawa}
\affiliation{Graduate School of Science, Hokkaido University, Sapporo 060-0810, Japan}
\author{Vladim{\'i}r~\surname{Sechovsk{\'y}}}
\affiliation{Faculty of Mathematics and Physics, Charles University, Ke Karlovu 5, 121 16 Prague 2, Czech Republic}
\author{Hiroshi Amitsuka}
\affiliation{Graduate School of Science, Hokkaido University, Sapporo 060-0810, Japan}

\date{\today}

\begin{abstract}
Single-crystalline UAu$_2$Si$_2$ has been grown by a floating-zone melting method, and its magnetic, thermal and transport properties have been investigated through measurements of magnetization, specific heat and electrical resistivity to reveal its peculiar magnetism.
It is shown that UAu$_2$Si$_2$ undergoes a second-order phase transition at \tm = 19 K, which had been believed to be ferromagnetic ordering in the literature, from a paramagnetic phase to an uncompensated antiferromagnetic phase with spontaneous magnetization along the tetragonal $c$-axis (the easy magnetization direction). 
The magnetic entropy analysis points to the itinerant character of 5f electrons in the magnetic ordered state of UAu$_2$Si$_2$ with large enhancement of the electronic specific heat coefficient of $\gamma$ $\sim$ 150 mJ/K$^2$mol at 2 K. It also reveals the relatively isotropic crystalline electric field effect of this compound, with contrast to the other relative isostructural compounds.
The observed magnetization curves strongly suggest that there is a parasitic ferromagnetic component developing below $\sim$ 50 K in high coercivity with the easy axis along the tetragonal $c$-axis.
The results are discussed in the context of evolution of magnetism within the entire family of isostructural U$T_2$Si$_2$ compounds.
\end{abstract}

\maketitle

\section{\label{sec:level1}Introduction}

Materials containing U elements have been studied intensively for several decades, revealed to exhibit various exotic physical properties such as variety of magnetic structures, heavy electrons and their superconductivity, coexistence between magnetism and superconductivity, higher-order multipole orderings, hidden order, and so on. In these materials, the 5f electrons of U atoms play a key role in the emergence of these interesting phenomena. However, a comprehensive framework for understanding of electronic and magnetic properties of 5f electron systems has not yet been formed. To solve it, there is the unavoidable problem rooted in the nature of 5f electrons themselves: ``How can we describe the dual nature of 5f electrons?" The localized/itinerant character of 5f electrons lie in between 3d and 4f electrons, which makes it difficult to set a proper model to approach experimental observations for 5f-electron systems. In order to contribute to this problem from the experimental aspect, it is valuable to provide a set of reliable data of well-characterized physical properties of various 5f-electron compounds. In particular, systematic studies of isostructural compounds are useful to simplify the problems and highlight the nature of phenomena.

The U$T_{2}X_{2}$ compounds of uranium, with transition-metal atoms at the \textit{T}-sites and silicon or germanium atoms at the \textit{X}-sites, provide good opportunities for such systematic studies. Above all, those with silicon atoms at the \textit{X}-sites have been investigated since the early period of research of actinide intermetallics. The U$T_2$Si$_2$ compounds form a variety of transition-metal elements which can occupy the \textit{T}-sites; it has been confirmed that there are thirteen stable compounds which contain each transition metal from Cr to Cu, from Ru to Pd, and from Os to Au in the 3d, 4d, 5d rows, respectively. Most of them except for systems of Fe, Os, and Ru (systems of Cr \cite{Matsuda03}, Mn \cite{Szytula88}, Co \cite{Chelmicki85}, Ni \cite{Chelmicki85, Lin91}, Cu \cite{Chelmicki85, Matsuda05, Honda06}, Rh \cite{Bak81,Palstra86}, Pd  \cite{Bak81,Honma98}, Ir \cite{Palstra86,Dirkmaat90,Verniere96}, Pt  \cite{Palstra86, Steeman90, Sullow08}, and Au \cite{Palstra86, Lin97}) order magnetically at transition temperatures ranging from $\sim$ 5 K to $\sim$ 100 K. Those of Fe and Os are no-ordering states with moderately enhanced Pauli paramagnetism  \cite{Szytula88, Palstra86}. The remaining one, URu$_2$Si$_2$, is well known to show the hidden order transition at 17.5 K \cite{Palstra85, Maple86, Schlabitz86}; its order parameter is still unidentified and has been studied intensively to this day. 

In contrast to piles of papers on URu$_2$Si$_2$, very few reports have been provided for UAu$_2$Si$_2$.
There have been only five reports about this compound since its discovery in 1986 by Palstra \textit{et al.} until the latest one in 1997 given by Lin \textit{et al}. \cite{Palstra86, Saran88, Rebelsky91, Torikachvili92, Lin97}.
All of them are about studies of polycrystalline samples, that is, no reference of single crystal growth has ever appeared, leaving low-temperature properties of the ordered state of this compound rather unclear.
According to most of the previous reports, UAu$_2$Si$_2$ crystallizes in the ThCr$_2$Si$_2$ type body-centered tetragonal (I4/mmm) structure adopted by the most other U$T_{2}X_{2}$ compounds.

So far strong sample dependence of physical properties of UAu$_2$Si$_2$ can be deduced from the existing literature, which makes things more complicated.
Lin $et$ $al$. investigated the annealing effects on this compound and pointed out differences in annealing conditions as the reason of the sample dependences \cite{Lin97}. 
They observed that as-cast samples exhibit ferromagnetic (FM) phase transitions at about 82 K and 20 K.
On the other hand, well annealed samples do not show any anomaly around 82 K in physical properties such as magnetization, specific heat and electrical resistivity.
Instead, another FM feature in magnetization appears at around 50 K followed by the 20-K transition similar to that in as-cast samples. Specific heat measurements, however reveals only the second-order phase transition at 20 K. 

Besides the consensus about the 20-K magnetic phase transition, other characteristics of magnetism in UAu$_2$Si$_2$ including magnetic anisotropy remained unexplored. 
In the present work, we succeeded in growing single crystals of UAu$_2$Si$_2$ by floating-zone melting method.
The crystals were investigated by detailed X-ray diffraction (XRD), specific heat, magnetization, and electrical resistivity measurements, which revealed features of peculiar anisotropic magnetism in this compound.

\section{\label{sec:level1}Experimental procedure}
\begin{figure}[h]
\centering
	\includegraphics[width=8.5cm]{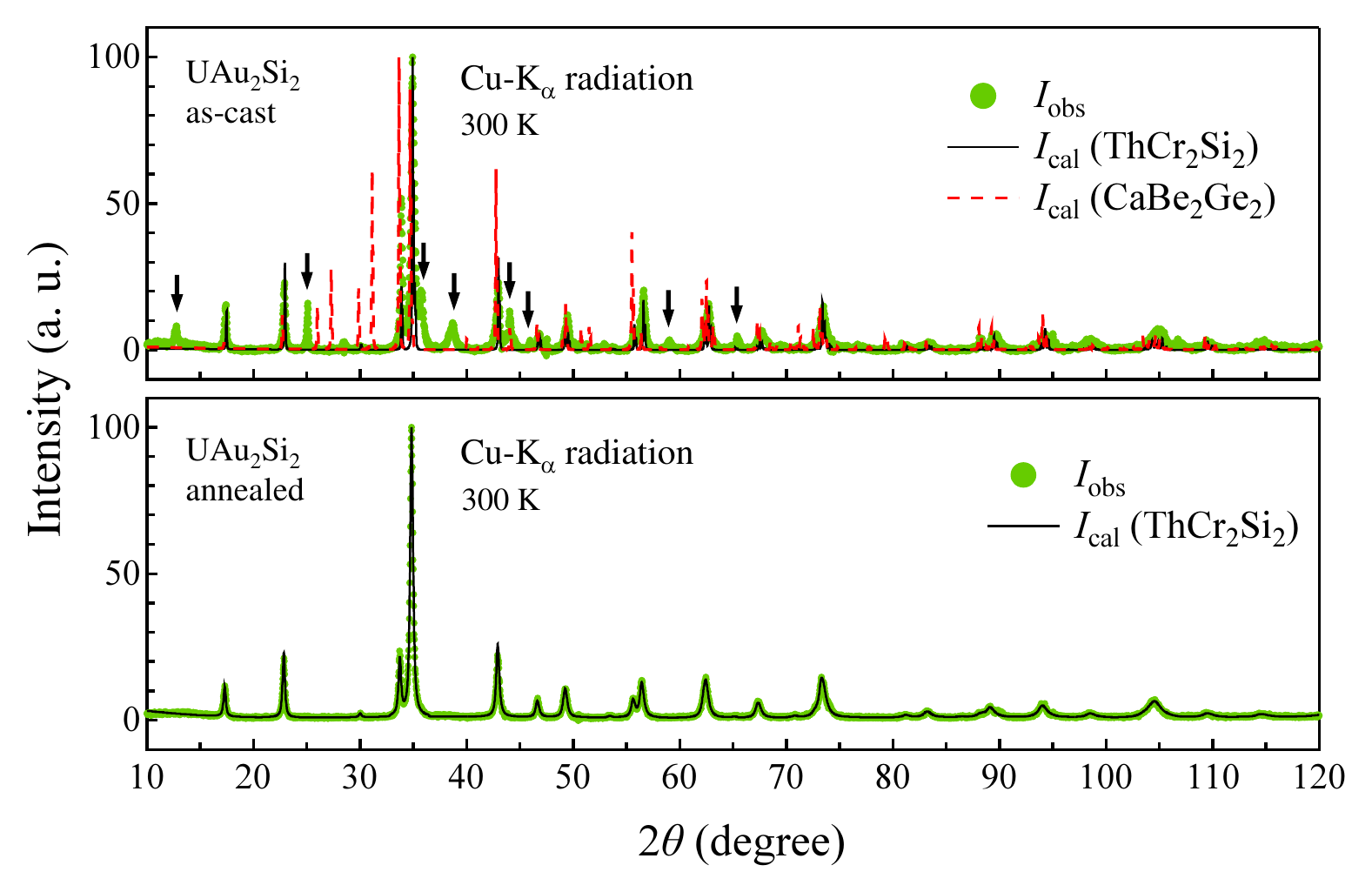}
	\caption{(Color online) X-ray powder diffraction (XRPD) patterns of as-cast (top) and annealed (bottom) polycrystals of UAu$_2$Si$_2$, displayed with simulated patterns assuming the ThCr$_2$Si$_2$ structure and CaBe$_2$Ge$_2$ structure. In the XRPD pattern of the as-cast sample, peaks that cannot be explained by either structures are indicated by arrows.}
	\label{XRD}
\end{figure}

Firstly, polycrystalline samples of UAu$_2$Si$_2$ and its non-5f counterpart ThAu$_2$Si$_2$ were synthesized by arc-melting in Ar atmosphere, with stoichiometric amount of the starting materials of U(99.9\%), Au(99.99\%) and Si(99.999\%).
X-ray powder diffraction (XRPD) patterns of the as-cast samples have some peaks which cannot be explained by either the ThCr$_2$Si$_2$ type structure or the CaBe$_2$Ge$_2$ type one, as shown in Fig. \ref{XRD}.
The intensity of the strongest unidentified peak is $\sim$10 \% of the main UAu$_2$Si$_2$ peak.
After annealing in vacuum at a temperature of 900$^{\circ}$C for 1 week, the unidentified peaks all vanished from XRPD patterns, and all remaining peaks were explained by the ThCr$_2$Si$_2$ type body-centered tetragonal structure.
The diffraction patterns measured at 293 K and 8 K were almost the same except for differences in peak positions corresponding to thermal expansion of the crystal lattice. 
The refined lattice parameters are listed in Table \ref{tab:lattice_parameters}.
Note that the thermal expansion of UAu$_2$Si$_2$ between 300 and 8 K is strongly anisotropic: the \textit{a}-axis shrinks four times as much as the \textit{c}-axis.
We also prepared polycrystalline ThAu$_2$Si$_2$ samples in the same manner as above, and observed similar annealing effects.

The single crystals of UAu$_2$Si$_2$ have been grown by floating zone melting method using an optical furnace by Crystal Systems Corporation, applied on the precursor polycrystalline rod annealed at 1000$^{\circ}$C for three days.
The rod after zone melting procedure was composed of many macroscopic crystallites of single crystal, which have different orientations from each other.
The size of the crystallites was about 1 mm.
For the obtained crystallites, a tetragonal structure with the lattice parameters \textit{a} = 4.213 \AA, \textit{c} = 10.31 \AA \ at room temperature was confirmed by single crystal XRD, and 4-fold rotational symmetric Laue patterns were observed. We also performed EDX analyses and confirmed that the stoichiometry ratio is approximately 1 : 2 : 2.

The magnetization was measured by a SQUID magnetometer of MPMS 7T in the temperature range from 2 to 350 K.
Where needed the magnetization measurements were extended to the higher magnetic fields up to 14 T by using the vibration sample magnetometer option of a PPMS 14T.
Specific heat was measured by the thermal relaxation technique in the temperature range from 5 to 200 K in the magnetic fields up to 9 T by a PPMS 9T.
Electrical resistivity was measured by the conventional four-probe method in the temperature range from 2 to 350 K in the magnetic fields of 0 T and 9 T by using the PPMS 9T.
The MPMS 7T and both the PPMS apparatuses were from Quantum Design Inc.

\renewcommand{\arraystretch}{1.5} 
\begin{table}
\caption{Lattice parameters and the atomic position of Si atoms of annealed polycrystalline UAu$_2$Si$_2$ obtained by X-ray powder diffraction with the Reitveld analyses using a software RIETAN-FP \cite{Izumi07}. The typical reliability factors are $R_{\rm wp} = 12 \%$, $R_{\rm F} = 3.8 \%$, $S = 1.8$.\\}
\begin{tabular}{cccc}
\hline
$T$ (K) & $a$ (\AA) & $c$ (\AA) & $z_{\rm Si}$\\
\hline
293       & 4.223(1)             & 10.290(1)          & 0.391(1)             \\
8           & 4.207(1)             & 10.280(1)         & 0.390(1)             \\
\hline
\end{tabular}
\label{tab:lattice_parameters}
\end{table}

\section{\label{sec:level1}Results}

\subsection{Specific heat}
\begin{figure}[h]
\centering
	\includegraphics[width=7.5cm]{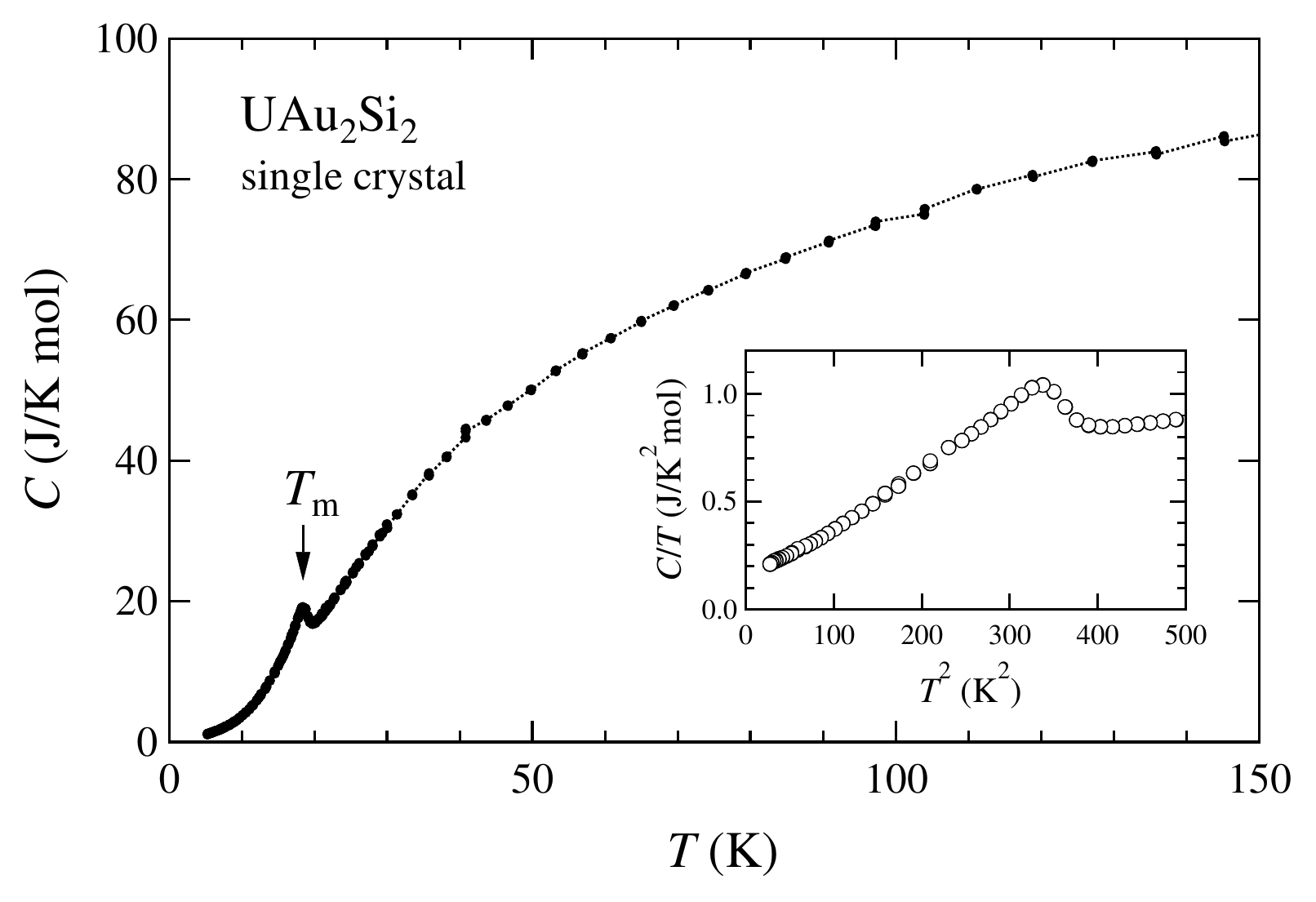}
	\caption{Temperature dependence of the specific heat of single-crystalline UAu$_2$Si$_2$. The inset shows the low-temperature-specific heat, $C$, divided by temperature, $T$, as a function of $T^2$.}
	\label{C_vs_T_0-300K}
\end{figure}

Figure \ref{C_vs_T_0-300K} shows the temperature dependence of the specific heat.
A distinct lambda anomaly was observed at 19 K, indicating an occurrence of second-order phase transition.
Now we label the transition temperature as \tm.
The phase transition is due to 5f electrons of uranium ions, because no anomaly was observed in the specific heat of polycrystalline ThAu$_2$Si$_2$ with no 5f electron, as shown in the later section.
The electronic specific-heat coefficient $\gamma$ estimated from linear extrapolation to $T$ = 0 of a $C/T$ versus $T^2$ plot (the inset of Fig. \ref{C_vs_T_0-300K}) shows a significantly large value of $\sim 150 \ {\rm mJ/K^{2} mol}$.

\begin{figure}[tbh]
	\centering
		\includegraphics[width=7cm]{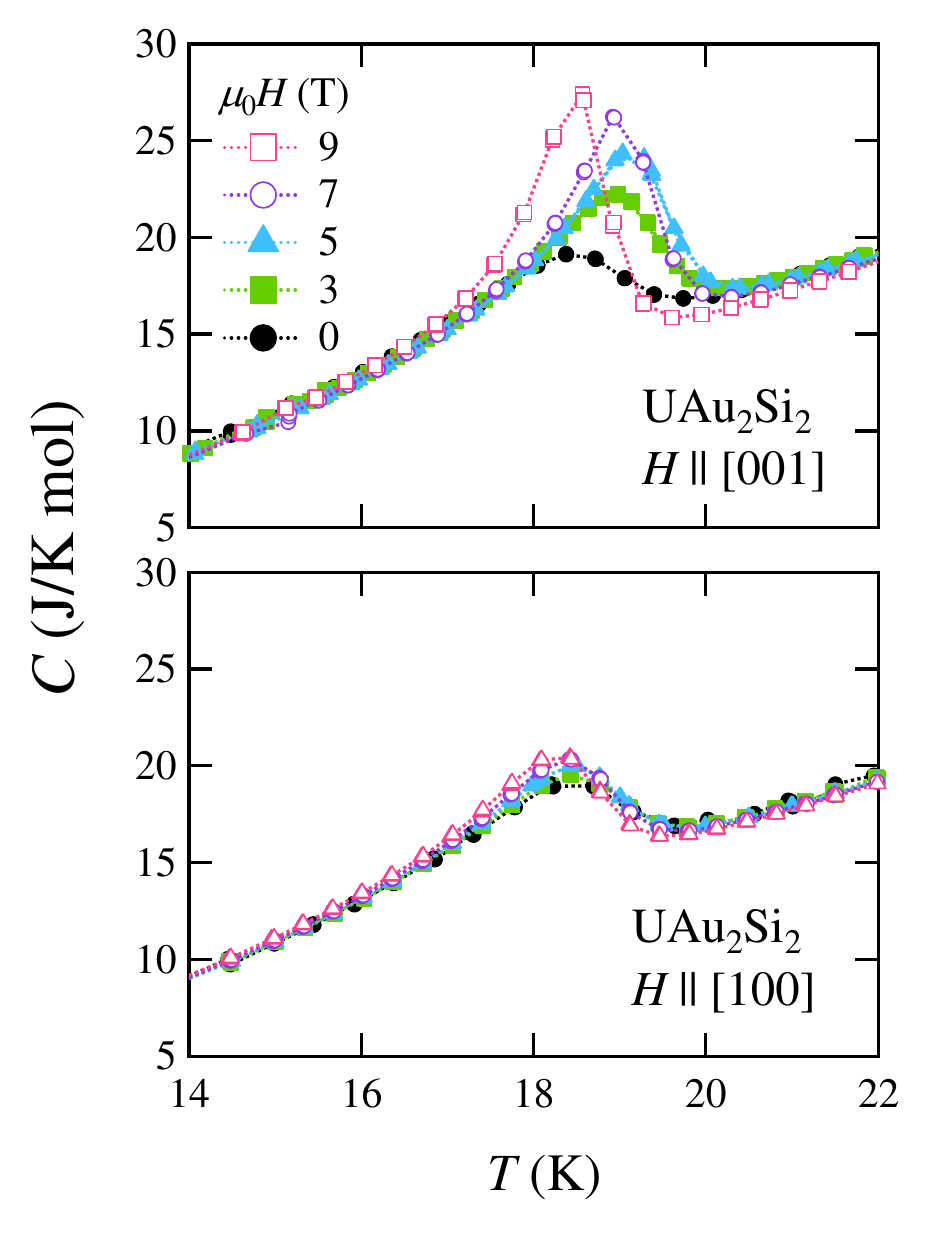}
	\caption{(Color online) Temperature dependence of the specific heat of single crystalline UAu$_2$Si$_2$ measured in magnetic fields along two crystallographic axes, [001] (upper panel) and [100] (lower panel).}
	\label{C_vs_T_12-26K}
\end{figure}

\begin{figure}[h]
	\centering
		\includegraphics[width=7.5cm]{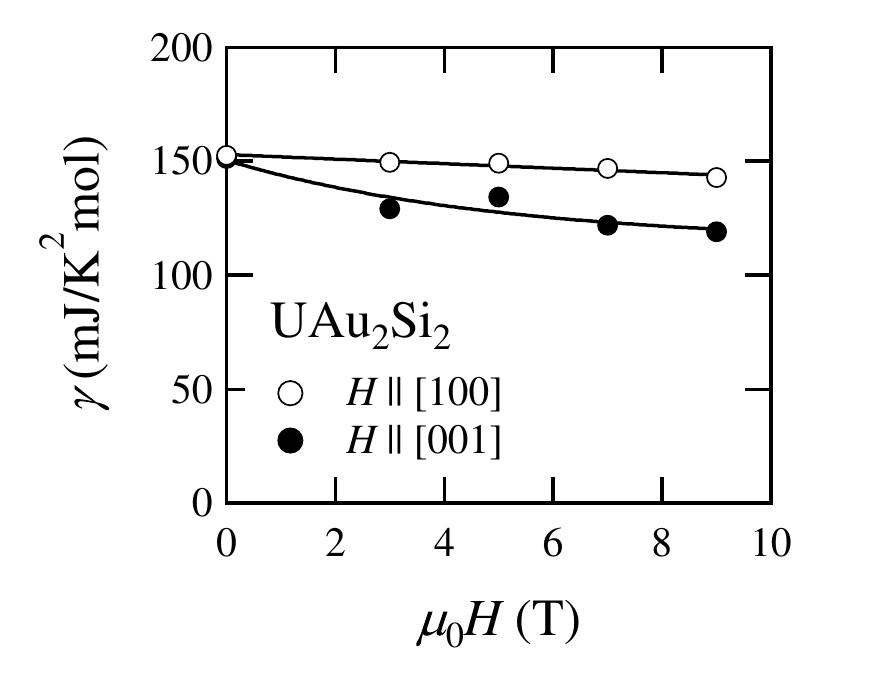}
	\caption{Magnetic-field dependence of the electronic specific heat coefficient $\gamma$ deduced from the linear extrapolation of the $C/T$ versus $T^2$ data to $T$ = 0 for each field for single-crystalline UAu$_2$Si$_2$. The solid curves are guides to the eye.}
	\label{g_vs_H}
\end{figure}

In magnetic fields, the specific heat around \tm behaves rather differently depending on a direction of the applied field as shown in Fig.\,\ref{C_vs_T_12-26K}.
Its temperature dependence shows a more pronounced peak anomaly at \tm by increasing the field along [001]; the peak becomes sharper and larger, meaning that the more entropy is released due to the phase transition. This is considerably different from the behavior that is expected for usual ferromagnetic systems, where a specific-heat peak associated with the FM transition becomes broader by applying magnetic fields.
In contrast, it does not show any significant change by applying the field along [100].  
This result suggests that the order below \tm becomes more stable by applying a magnetic field only along the [001] axis, implying strongly anisotropic magnetic interactions in UAu$_2$Si$_2$.
Figure \ref{g_vs_H} shows the magnetic field dependence of the $\gamma$ values estimated from the linear extrapolation of the $C/T$ versus $T^2$ data to $T$ = 0 for each field below 7 K. The $\gamma$ value is reduced by increasing the field along [001]; it decreases about 20 percent in the field of 9 T, corresponding to the simultaneous enhancement of the entropy release at the transition temperature. 


\subsection{Electrical resistivity}
\begin{figure}[h]
		\includegraphics[width=7.5cm]{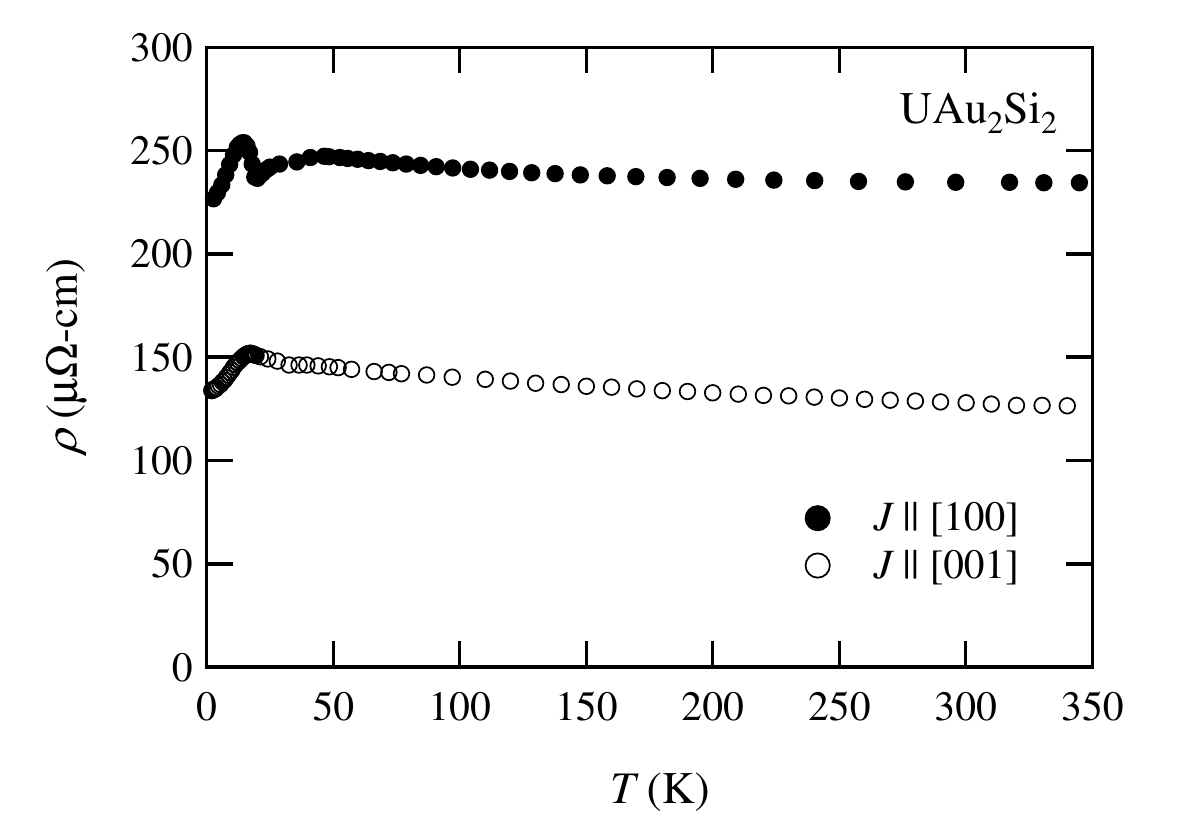}
		\caption{Temperature dependence of the electrical resistivity of UAu$_2$Si$_2$ for electric currents along [100] and [001].}
		\label{rho_vs_T}
\end{figure}

The temperature dependences of electrical resistivity for the [100] and [001] directions of electric currents $J$ are shown in Fig. \ref{rho_vs_T}.
The observed behavior is far from the typical metallic ones, similarly to numerous other U intermetallics \cite{Sechovsky98}: the resistivity at high temperatures increases with decreasing temperature for both the crystallographic axes of [100] and [001].
The increase of the [100] resistivity with decreasing temperature terminates around 40 K, which is followed by gradual decrease with decreasing temperature down to \tm, whereas the [001] resistivity continues increasing.

\begin{figure}[h]
	\centering
		\includegraphics[width=8cm]{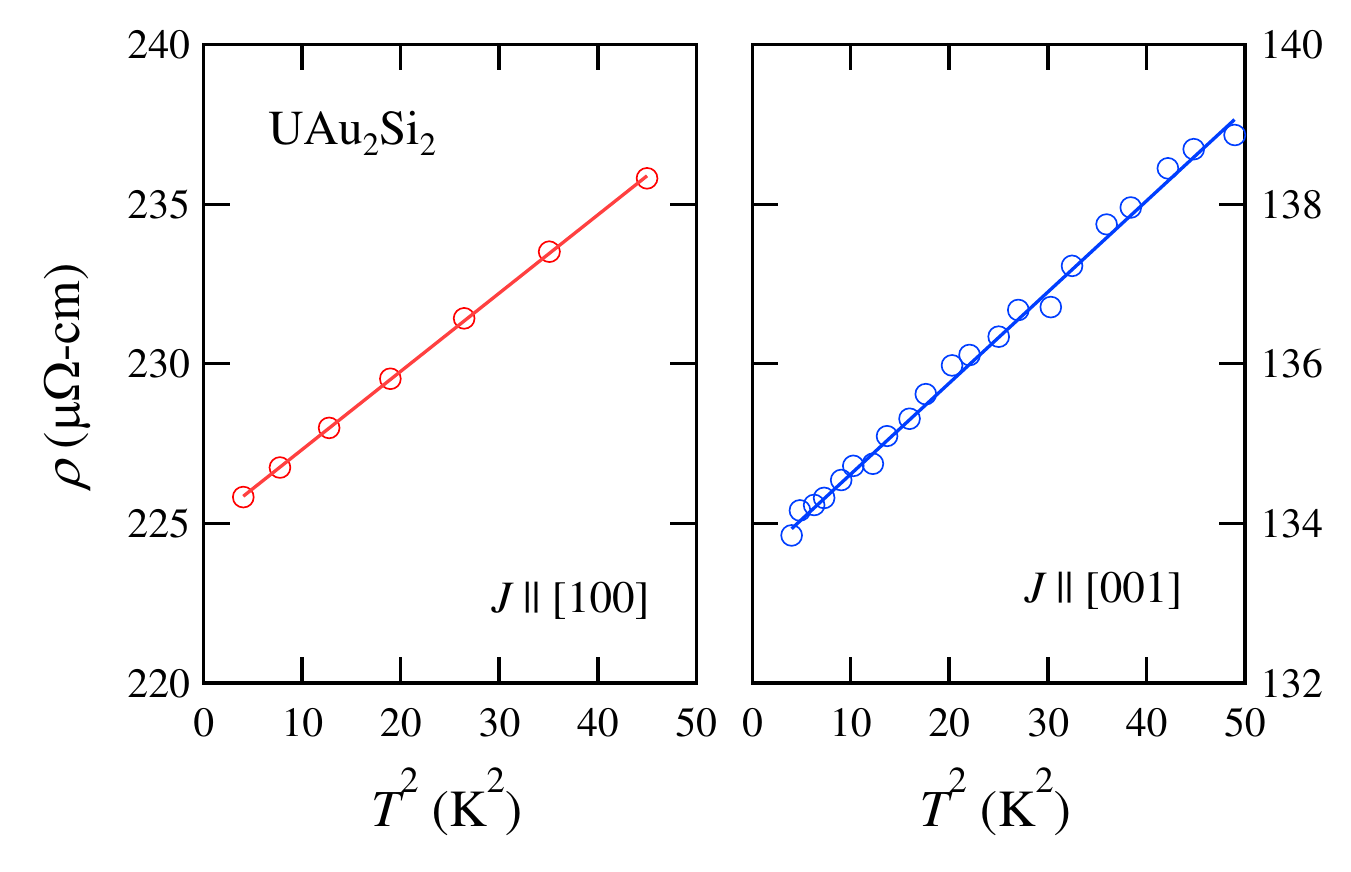}
	\caption{(Color online) ${\rho}$ vs. $T^2$ plot below $\sim$7 K in current directions of [001] (left) and [100] (right). Solid lines are fitting curves using the function described in the text.}
	\label{rho_vs_Tsq}
\end{figure}

At \tm, the resistivity shows an upturn in both directions of the current.
(The upturn in the current along [001] is very subtle, but it does exist.)
It may be associated with the opening of a gap on the Fermi surface due to the phase transition.
This kind of anomaly in electrical resistivity suggestive of the reduction of the carrier number is also observed in other uranium 1-2-2 compounds, such as URu$_2$Si$_2$ \cite{McElfresh87}, UCo$_2$Si$_2$ \cite{Mihalik07}, and UNi$_2$Ge$_2$ \cite{Ning92}.
At lower temperatures the resistivity finally decreases with decreasing temperature. 
The temperature dependence of the resistivity below $T_{\rm m}$ cannot be fitted by a function which contains a term of ${\rm exp}(-\Delta / T)$ assuming an opening of a gap of $\Delta$ on the Fermi surface. Such a description has given fairly good agreements with the data for URu$_2$Si$_2$ \cite{McElfresh87} and UNi$_2$Ge$_2$ \cite{Ning92}.
Instead, the data on UAu$_2$Si$_2$ simply shows the $T^2$ dependence below 7 K as shown in Fig. \ref{rho_vs_Tsq}.
The best fit using the function of ${\rho}(T) = {\rho}_0 + AT^2$ gives the coefficients $A$ with strongly enhanced values: $A \sim 0.24$ ${\mu}{\Omega}{\rm cm} {\rm K}^{-2}$ for $J$ $\parallel$ [100] and $A \sim 0.12$ ${\mu}{\Omega}{\rm cm} K^{-2}$ for $J$ $\parallel$ [001].
At the lowest temperature, 2 K, the high residual resistivity is observed, probably reflecting one or both of contributions due to large magnetic scattering and crystal defects. 

\subsection{Magnetization}
\label{sec:result-Magnetization}
\begin{figure}[h]
	\centering
		\includegraphics[width=7cm]{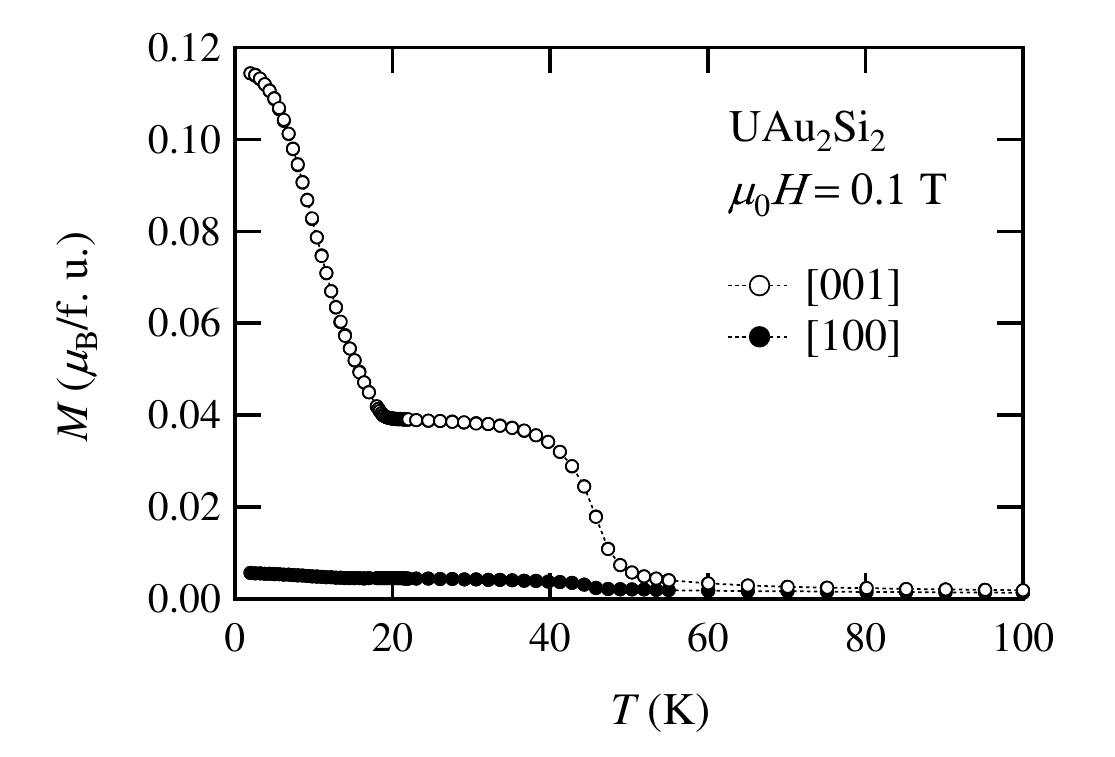}
	\caption{Temperature dependence of the magnetization of UAu$_2$Si$_2$ measured in magnetic field of 0.1 T.}
	\label{M_vs_T_01T}
\end{figure}

\begin{figure}[h]
	\centering
		\includegraphics[width=7cm]{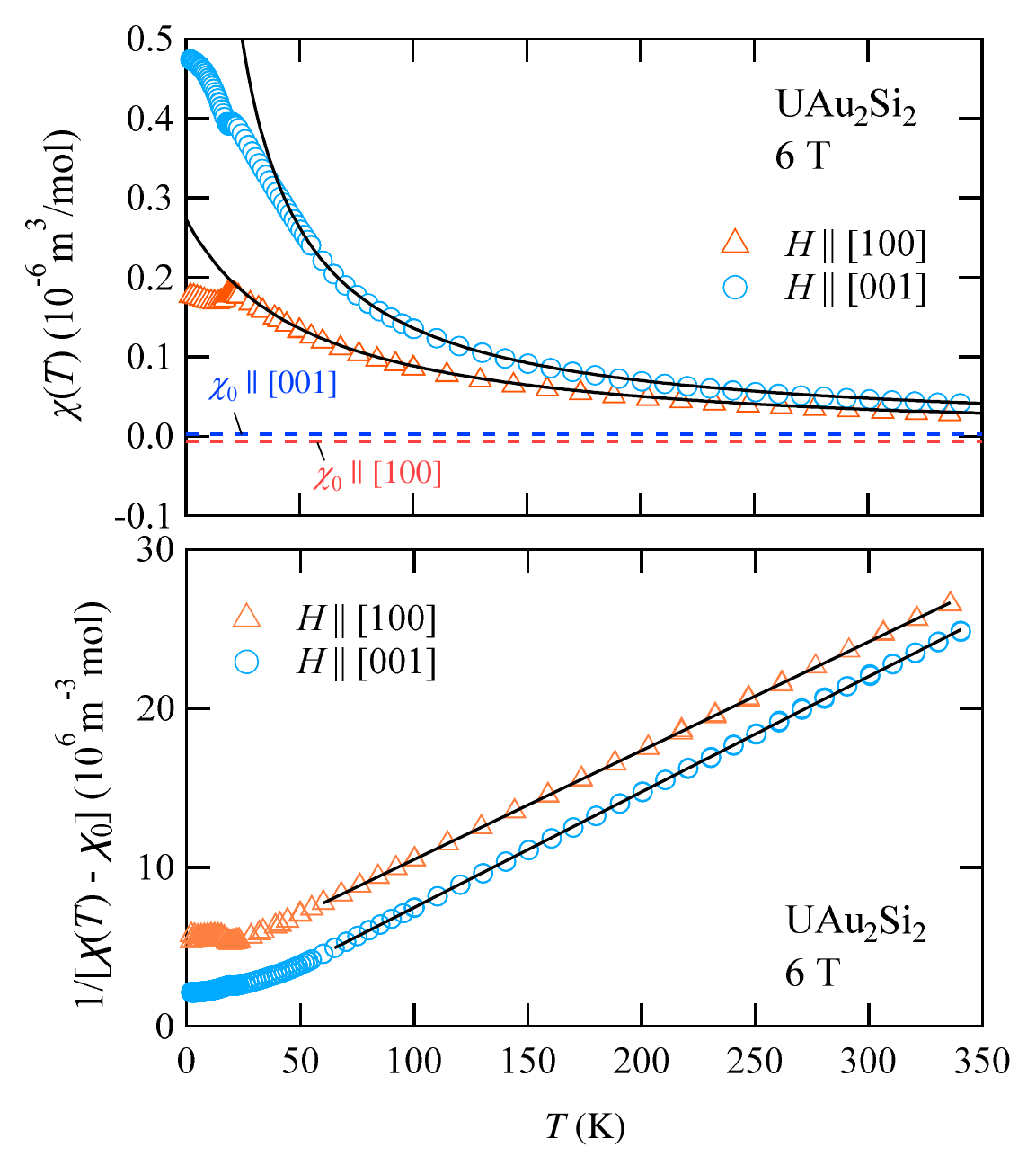}
	\caption{(Color online) Temperature dependence of the magnetic susceptibility (top) and inverse susceptibility (bottom) from which the constant contribution is subtracted by using Eq. (\ref{eq_CW}) described in the text (bottom). Solid lines indicate the best fits to the data above 60 K using Eq. (\ref{eq_CW}). In the top panel, the constant susceptibility components are displayed by broken lines.}
	\label{chiinv_vs_T}
\end{figure}

The considerable differences between the corresponding [100] and [001] magnetization  at low temperature document strong magnetocrystalline anisotropy of UAu$_2$Si$_2$ as presented in Fig. \ref{M_vs_T_01T}. The easy axis is the $c$-axis as in most of the other uranium 1-2-2 systems. 
As we can see in Fig. \ref{chiinv_vs_T}, in a higher temperature region above $\sim$ 60 K the paramagnetic susceptibility for the both axes follows the modified Curie-Weiss's law:
\begin{equation}
\chi (T) = \frac{C}{T-{\it {\Theta}}_{\rm W}} + \chi_{\rm 0},
\label{eq_CW}
\end{equation}
where $\chi_0$ is a temperature independent term which is considered to include the contributions of Pauli paramagnetism of conduction electrons, diamagnetism of core electrons, and a Van-Vleck term of 5f-electrons.
For both the axes, the fitting analyses give small values of $\chi_{\rm 0}$ in an order of 10$^{-9}$ m$^3$/mol.
The effective magnetic moment and the Weiss temperature are estimated as $\mu_{\rm eff} =  3.05(10)$ {\mub}/U and ${\it {\Theta}}_{\rm W} = -52 \pm 10$ K for $H$ $\parallel$ [100], and  $\mu_{\rm eff} =  2.96(10)$ {\mub}/U and ${\it {\Theta}}_{\rm W} = -3 \pm 10$ K for $H$ $\parallel$ [001].
The negative ${\it {\Theta}}_{\rm W}$ values indicate the presence of antiferromagnetic exchange interactions in UAu$_2$Si$_2$. The 49 K difference between the ${\it {\Theta}}_{\rm W}$ values for the [100] and [001] directions is considered to produce a major part of  the magnetic anisotropy in the paramagnetic state, since the effective $g$-factor is estimated to be nearly isotropic. 
This feature is quite a contrast to many of other U$T_2$Si$_2$ compounds which show strong Ising-type uniaxial anisotropy.
The derived values of effective moment per uranium ion of approximately 3 Bohr magnetons for both the principal directions are around 80 percent as large as the U$^{3+}$ and U$^{4+}$ free ion values (3.62 and 3.58 \mub, respectively).
This discrepancy suggests delocalization of the uranium 5f electrons.

At lower temperature below around 50 K, the magnetization shows strongly anisotropic behaviors, not only the magnitude but also its temperature dependence.
Around 50 K, an upturn was observed along the both crystallographic axes in the \textit{M-T} curves measured in very low fields. 
The magnitude of the upturn is very small and anisotropic; it is around 0.03 \mub per uranium ion along the [001] and it is much less than 0.01 \mub along the [100]. We cannot straightforwardly identify this FM anomaly as an onset of the bulk phase transition because no anomaly is observed in the specific heat at around 50 K.
On the other hand, the electrical resistivity shows a maximum at around 50 K for the direction of electric currents along [100]. However, this anomaly is not sharp, and should certainly be irrelevant to the development of ferromagnetism.
The possible origin of this FM component will be discussed in Sec.\,\ref{sec:discussion-50K}.
Another upturn anomaly at around 20 K is obviously caused by the phase transition at \tm, seemingly indicating a ferromagnetically ordered state below this temperature as reported by previous studies. The magnetically ordered state is highly anisotropic, with small cusp anomaly along [100] direction at \tm. 

\begin{figure}[h]
	\centering
		\includegraphics[width=7cm]{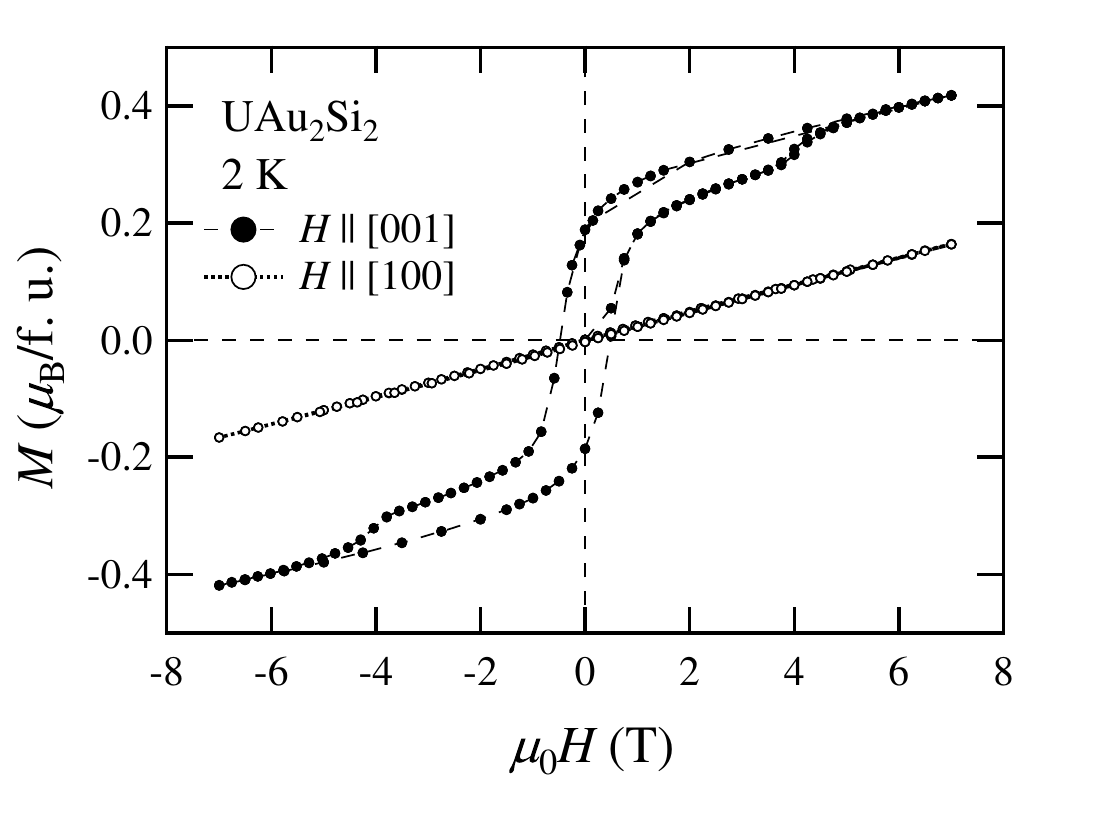}
	\caption{The magnetization processes in single-crystalline UAu$_2$Si$_2$ at 2 K.}
	\label{M_vs_H}
\end{figure}

\begin{figure}[h]
	\centering
		\includegraphics[width=7cm]{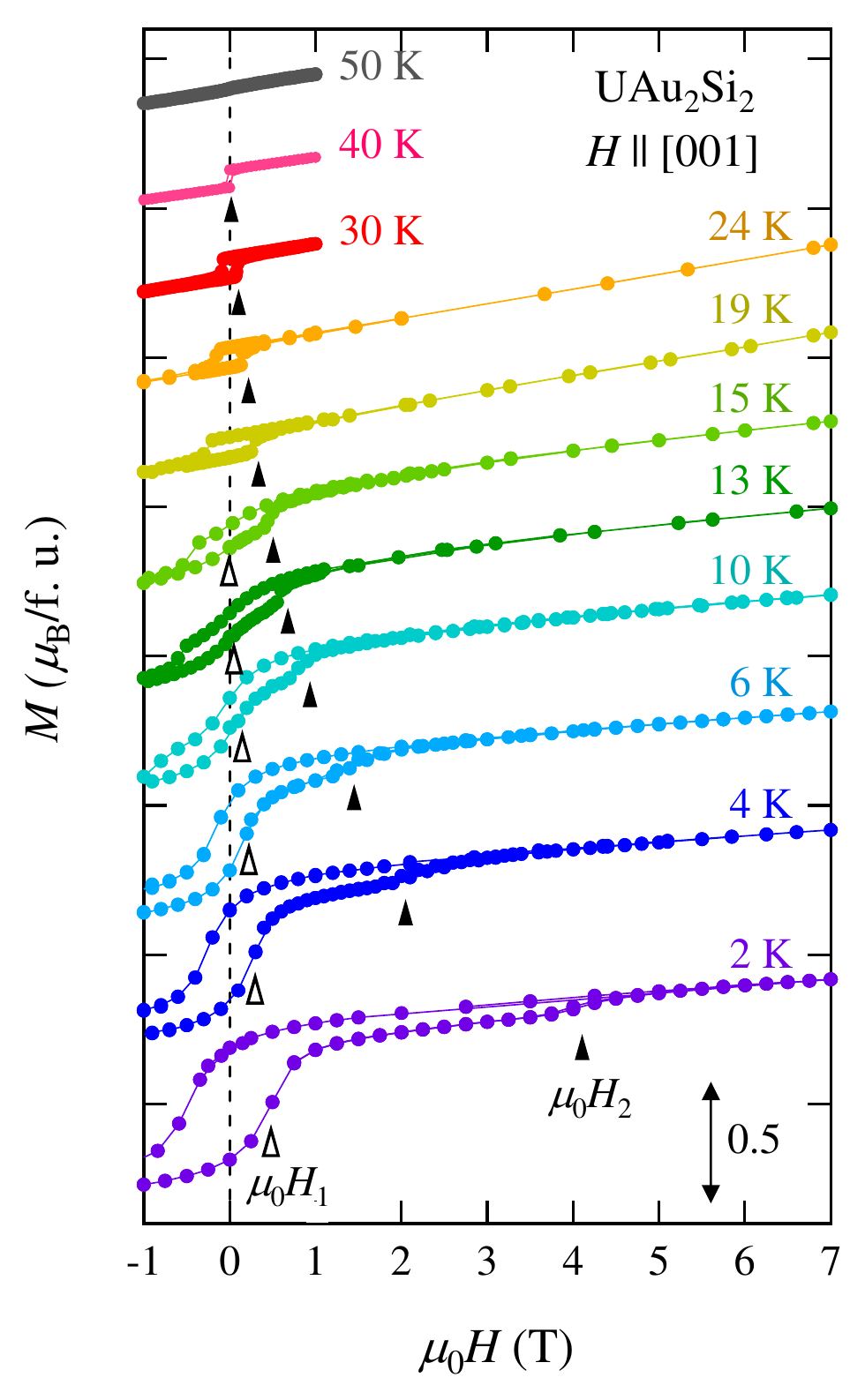}
	\caption{(Color online) Temperature dependence of the magnetization curves in single-crystalline UAu$_2$Si$_2$ measured in magnetic field parallel to the tetragonal $c$-axis. Each curve is shifted vertically. }
	\label{M_vs_H_all}
\end{figure}

\begin{figure}[h]
	\centering
		\includegraphics[width=7cm]{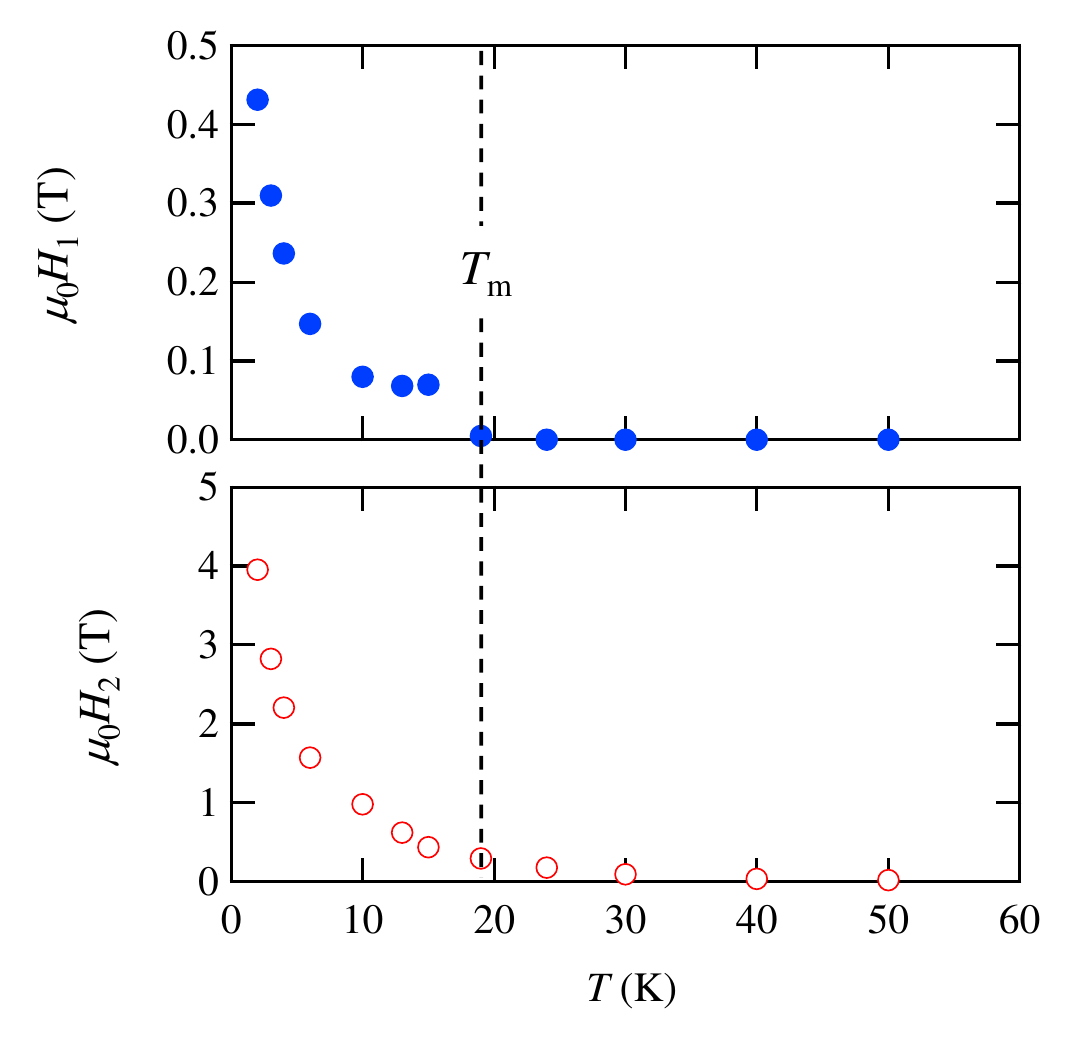}
	\caption{(Color online) Temperature dependence of $H_1$ (top) and $H_2$ (bottom), at which fields the step-like anomaly is observed in the magnetization curves.}
	\label{H1H2_vs_T}
\end{figure}

These FM components are also confirmed in a distinct hysteresis loop of the magnetization process in magnetic field along [001] displayed in Fig.\,\ref{M_vs_H}.
It is quite a contrast to the magnetization along [100], which is simply proportional to the applied magnetic field at all measurement temperature points ranging from 2 K to 60 K.
A striking feature of the hysteresis loop is its complex shape with step-like structures at 2 K.
Here we define two magnetic-field points for each step, $H_{1}$ and $H_{2}$, as the inflection points as shown in Fig.\,\ref{M_vs_H}.
Figure \ref{M_vs_H_all} shows the temperature dependence of the hysteresis observed at some temperature points from 2 K to 60 K.
Both $H_{1}$ and $H_{2}$ decrease by rising temperature. 
Fig. \ref{H1H2_vs_T} shows the temperature dependence of $H_{1}$ and $H_{2}$, which were obtained from an analysis using two-component fitting described in Sec. \ref{sec:discussion-50K}.
What should be noted here is the temperature at which they vanish. $H_{1}$ goes to zero at around 19 K, namely \tm. This indicates that it attributes to the magnetic order occurring below \tm.
On the other hand $H_{2}$ survives even above \tm, and eventually goes up to zero by increasing temperature up to 50 K, corresponding to the onset of the small FM component observed in the $M$-$T$ curve. The detailed analyses will be given in Sec. \ref{sec:discussion-50K}.


\begin{figure}[tb]
	\centering
		\includegraphics[width=7cm]{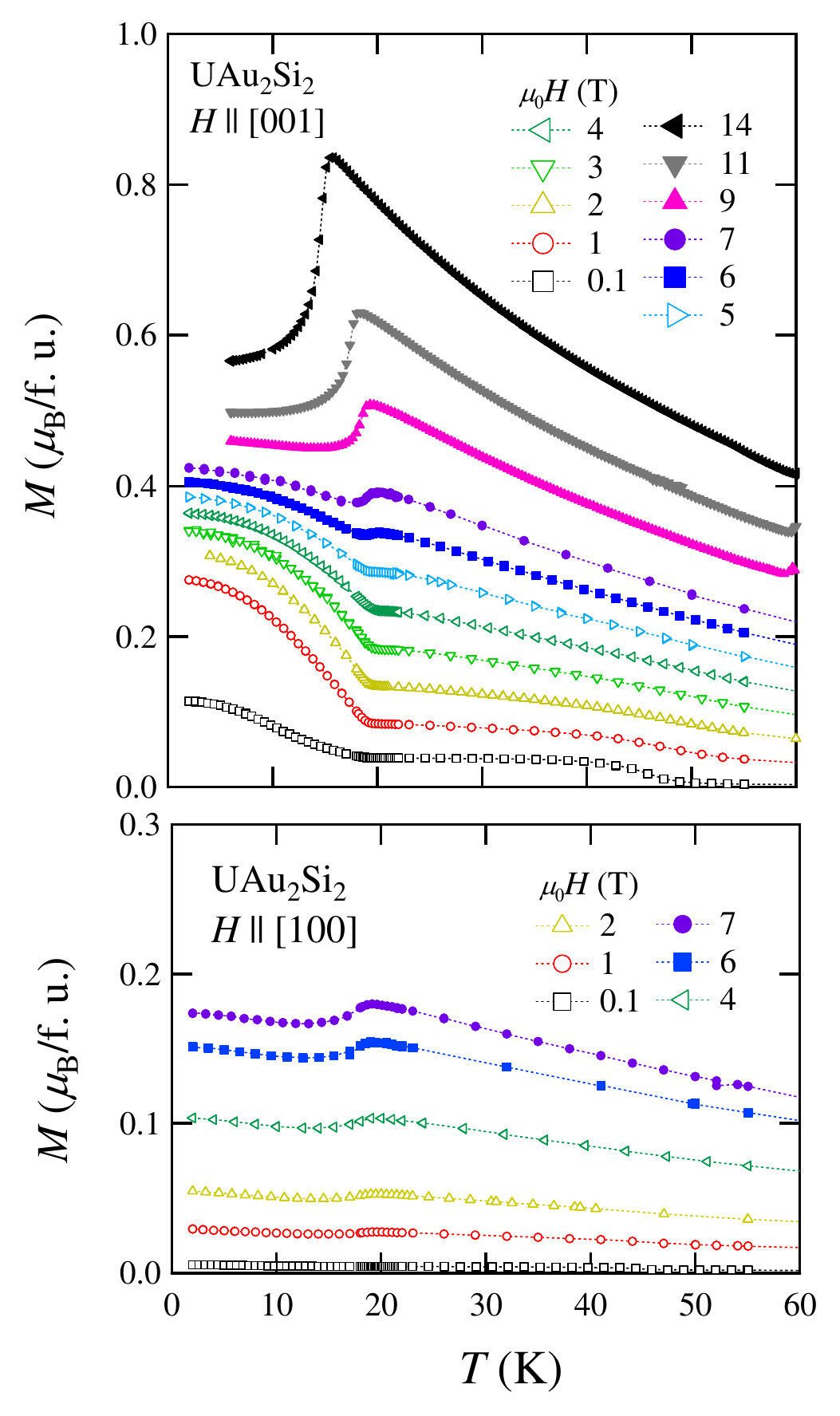}
	\caption{(Color online) Temperature dependence of magnetization in UAu$_2$Si$_2$ in magnetic fields applied along the [001] (upper panel) and [100] (lower panel) axis. The closed and open circles are field-cool (FC) and zero-field-cool (ZFC) data, respectively.}
	\label{M_vs_T_field}
\end{figure}

The temperature dependence of magnetization shows a remarkable change by increasing applied magnetic field at temperatures around \tm, as shown in Fig. \ref{M_vs_T_field}.
We found that the magnetization along the [001] direction shows a cusp anomaly at \tm in magnetic fields above 5 T while the magnetization along [100] simply increases by applying field.
The FM like upturn is gradually suppressed by increasing the field above 2 T, resulting in disappearing (or just becoming invisible) above 7 T.
Simultaneously, the cusp anomaly is dramatically enhanced.
This, together with the magnetic-field-enhanced anomaly of the specific heat around \tm (see Fig.\,\ref{C_vs_T_12-26K}), strongly suggests that the magnetically ordered state of UAu$_2$Si$_2$ below \tm is not simply FM but including an antiferromagnetic (AFM) component in its magnetic structure.

\begin{figure}[h]
	\centering
		\includegraphics[width=7cm]{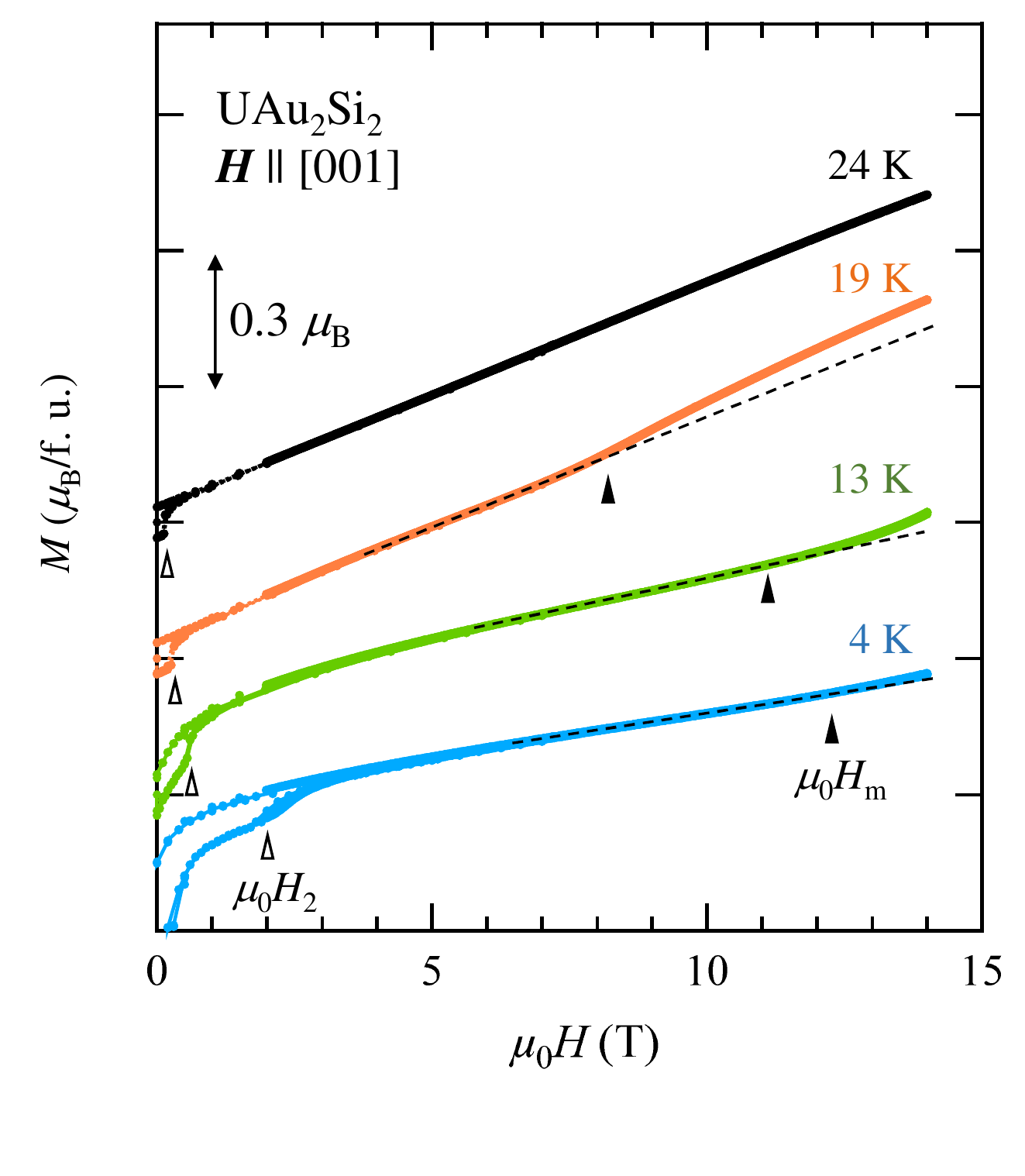}
	\caption{(Color online) Magnetization processes in UAu$_2$Si$_2$ up to 14 T along the [001] direction. The broken lines are guides to the eye.}
	\label{M_vs_H_highfield}
\end{figure}

We further investigated the magnetization process by extending the measured field range up to 14 T in the [001] direction.
The results are shown in Fig.\,\ref{M_vs_H_highfield}.
Besides the $H_2$ anomaly, we found that the magnetization curve bends upwards in a high-field region below \tm.
Although the overall features of the magnetization curves are unclear in this field range particularly at low temperature, we here simply define $H_{\rm \scalebox{0.5}{m}}$ as the field at which the magnetization starts to deviate from a linear field dependence. The roughly estimated $H_{\rm \scalebox{0.5}{m}}$ values are indicated by arrows in Fig. \ref{M_vs_H_highfield} and plotted in the $H$-$T$ phase diagram (Fig. \ref{HT_phase_line}) with error bars representing the ambiguity of estimation.
 This might be an implication that $H_{\rm \scalebox{0.5}{m}}$ anomaly does not correspond to a phase transition. Nevertheless, we consider that some sort of properties of the magnetic ordered state gradually changes in quality roughly above $H_{\rm \scalebox{0.5}{m}}$.
We suggest that the origin of the $H_{\rm \scalebox{0.5}{m}}$ anomaly is intrinsically different from that of the $H_2$, because no hysteresis has been observed around $H_{\rm \scalebox{0.5}{m}}$ unlike around $H_2$.
$H_{\rm \scalebox{0.5}{m}}$ increases by decreasing temperature down to 4 K, at which $H_{\rm \scalebox{0.5}{m}}$ reaches approximately 12 T.
The magnetization does not reach saturation at 14 T.
No $H_{\rm \scalebox{0.5}{m}}$-anomaly was observed at temperatures above \tm as manifested by the linear magnetic-field dependence of the magnetization at 24 K in Fig. \ref{M_vs_H_highfield}.

\subsection{Magnetic field-temperature phase diagram}

\begin{figure}[h]
		\includegraphics[width=7cm]{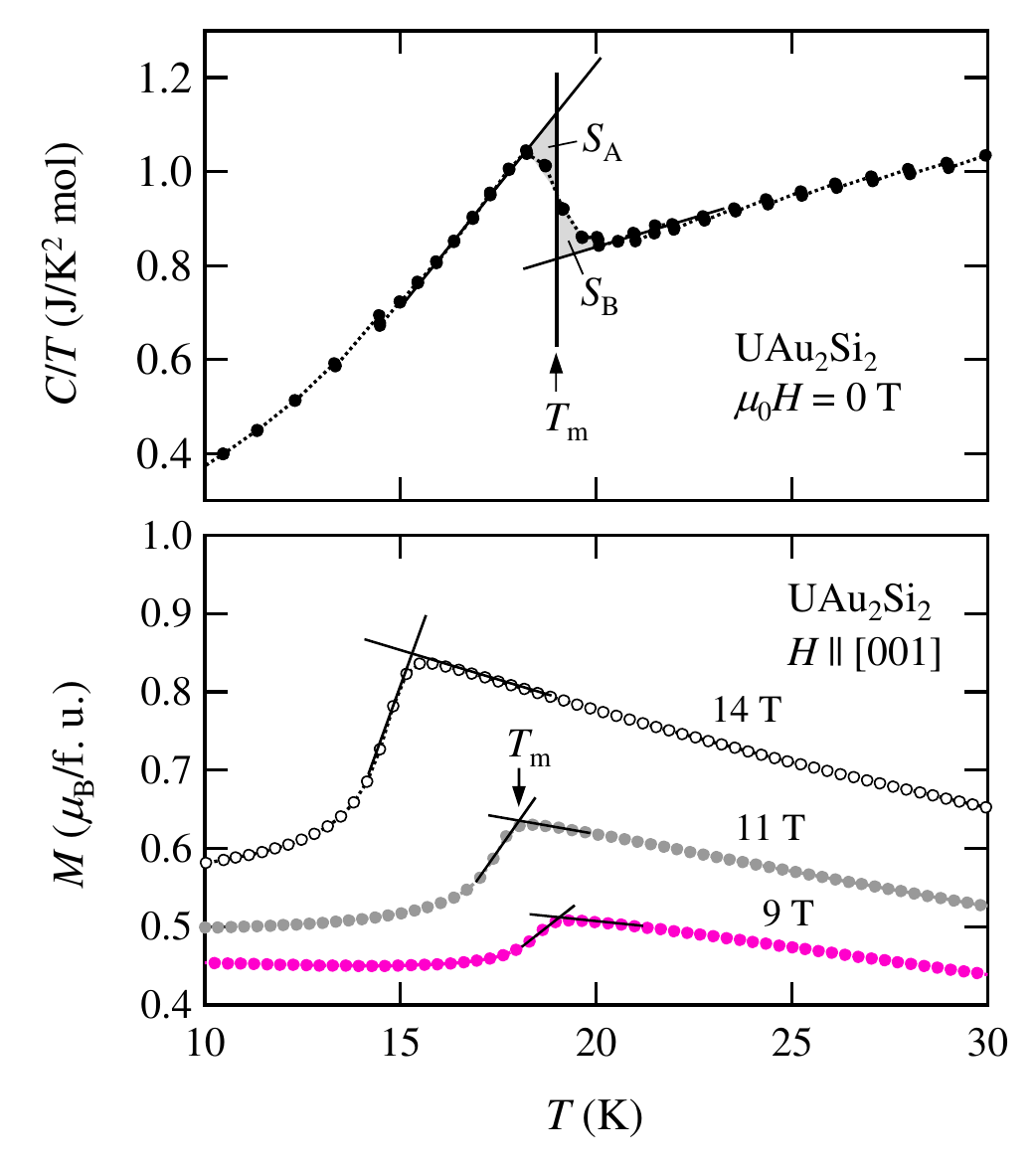}
	\caption{(Color online) Illustrations of how to determine the phase transition temperature, $T_{\rm m}$, by using the measured data of specific heat (upper panel) and magnetization (lower panel). 
From the specific heat, $T_{\rm m}$ was determined so that the condition $S_{\rm A} = S_{\rm B}$ is fulfilled. Here $S_{\rm A} $ and $S_{\rm B}$ are the area defined by extrapolation (the solid lines) of the data (the shaded area). 
From the magnetization, the cusp positions were used to determine $T_{\rm m}$.
}
	\label{det_Tm}
\end{figure}

\begin{figure}[h]
		\includegraphics[width=7.5cm]{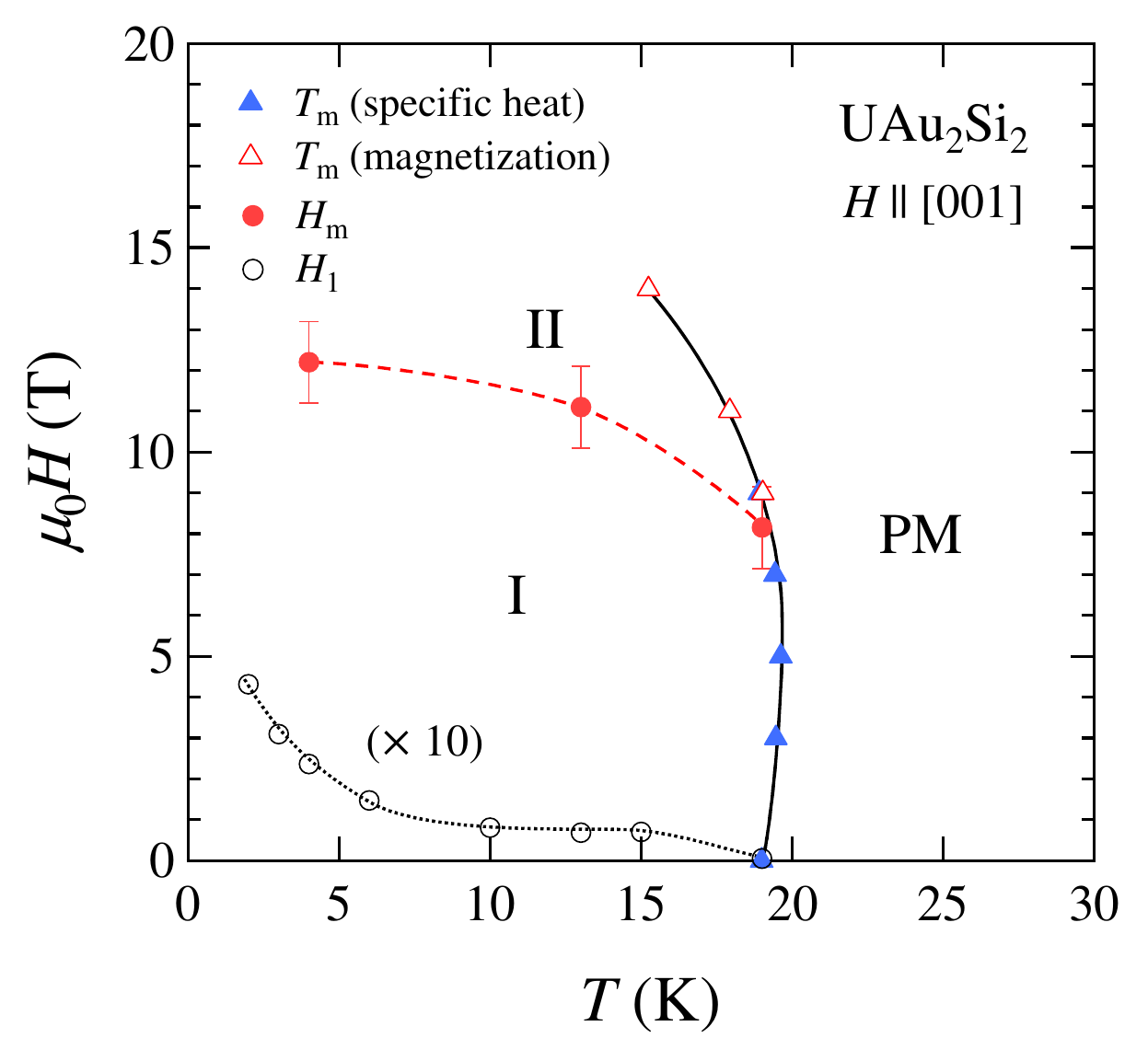}
	\caption{(Color online) \textit{H-T} phase diagram of UAu$_2$Si$_2$ for magnetic fields applied along the [001] direction. The full, broken and dotted lines are guides to the eye.}
	\label{HT_phase_line}
\end{figure}

We constructed a magnetic field-temperature (\textit{H-T}) phase diagram of UAu$_2$Si$_2$ for the applied field along the [001] direction in Fig.\,\ref{HT_phase_line}. \tm was determined from the temperature dependences of specific heat and magnetization in a magnetic-field range from 0 to 9 T and 9 to 14 T, respectively in the ways illustrated in Fig. \ref{det_Tm}. In the specific heat, \tm was determined to be the temperature that balances the entropy released at the phase transition. 
In the magnetization, on the other hand, we determined \tm more simply to be the temperature at which the magnetization shows a cusp. $T_1$ are also plotted in the phase diagram, as the intrinsic low-temperature character of the magnetic ordered state of UAu$_2$Si$_2$.


It is found that \tm goes up as the magnetic field increases in a low field range below $\sim$ 5 T. This provides a remarkable contrast to the behavior of many other AFM compounds. 
In usual AFM compounds, N\'{e}el temperature decreases when the strength of applied magnetic field is increased as expected theoretically in the classical molecular field approximation  in the early researches\cite{Burr64,Schmidt67}.
In fact, the isostructural relative U antiferromagnets, UPd$_2$Si$_2$\cite{Honma98}, UCr$_2$Si$_2$\cite{Matsuda03}, and UPt$_2$Si$_2$\cite{Schulze12} exhibit monotonous decrease in the N\'{e}el temperature with increasing magnetic fields.
On the other hand, a well-known quasi two-dimensional heavy fermion compound CeRhIn$_5$ has been reported to exihibit stabilization of AFM order with increasing magnetic field \cite{Sakurazawa05, Knebel11}. A two-dimensional Hubbard model taking into account the quantum fluctuation has been proposed to explain it\cite{Sakurazawa05}.
The enhancement of AFM order in a quasi-two dimensional antiferromagnet Cu(pz)$_2$(CIO$_4$)$_2$ has also been investigated in terms of a low-dimensional frustrated Heisenberg system \cite{Schmidt13, Fortune14}.
UAu$_2$Si$_2$ shows no implication of low-dimensional property, but at least it might be a frustrated magnetic system, as  showing both FM and AFM features in the ordered phase.  
In the high magnetic fields above 5 T, \tm decreases as the field is increased.
In order to see where the phase-boundary line towards, the experiments in higher magnetic fields are necessary.

Since $H_{\rm \scalebox{0.5}{m}}$ cannot be  identified as a phase boundary at the present stage, we refer to the two areas in the phase diagram divided by $H_{\rm \scalebox{0.5}{m}}$ as Area I and Area II.
The onset of $H_{\rm \scalebox{0.5}{m}}$ seems to roughly correspond to the lowest magnetic field where the cusp anomaly of the $c$-axis magnetization appears.
Another notable feature of the diagram is that the ordered state of Area I is stabilized by applying a magnetic field, whereas that of Area II is destabilized by increasing field.
These facts suggest that the nature of the magnetically ordered state is different between the Areas I and II.

\subsection{Magnetic-entropy analysis}
\begin{figure}[hbt]
	\centering
		\includegraphics[width=7cm]{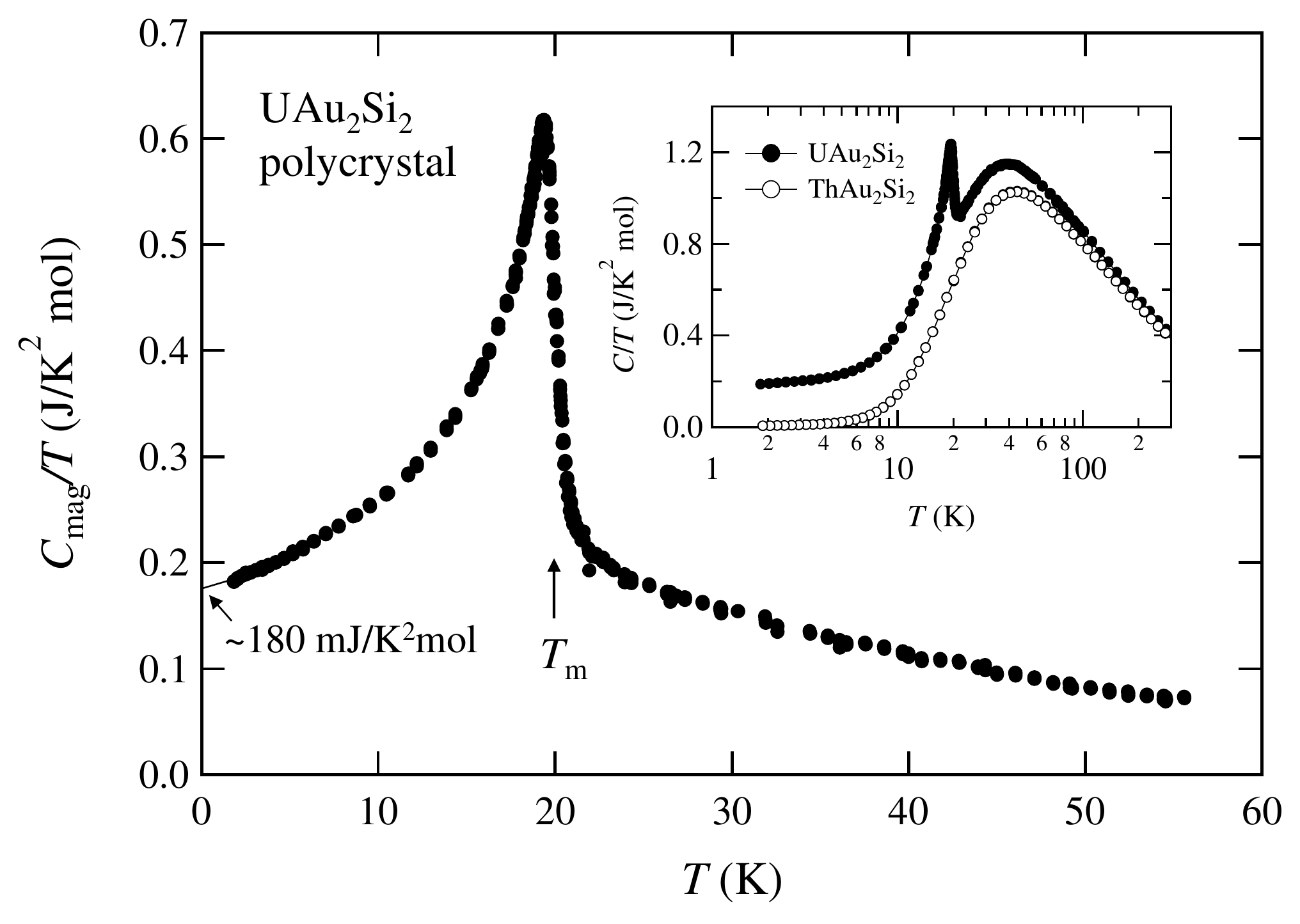}
	\caption{5f electronic contribution to the specific heat divided by temperature of UAu$_2$Si$_2$, derived by subtracting the ThAu$_2$Si$_2$ data (open circles in the inset) from the UAu$_2$Si$_2$ data (closed circles in the inset). The solid line is an extrapolation of the measured specific heat down to $T$ = 0 for a rough estimation of the $\gamma$-value.}
	\label{Cmag_vs_T}
\end{figure}

For investigation of the magnetic entropy of 5f electrons in UAu$_2$Si$_2$, we measured the specific heat of polycrystalline samples of ThAu$_2$Si$_2$ and UAu$_2$Si$_2$.
Figure \ref{Cmag_vs_T} shows the temperature dependence of 5f-electronic contribution of the specific heat divided by temperature, $C_\textrm{\scalebox{0.6}{mag}}/T$, obtained by subtracting the specific heat of ThAu$_2$Si$_2$, which is also shown in the inset of the figure.
At higher temperature above \tm, $C_\textrm{\scalebox{0.6}{mag}}/T$ increases monotonously with decreasing temperature, suggesting that the entropy release of 5f-electrons through c-f hybridization begins even in the paramagnetic state.  
Then it shows a distinct peak anomaly at $C_\textrm{\scalebox{0.6}{mag}}/T$, which indicates a phase transition caused by 5f electrons.
The extrapolated $\gamma$ value is roughly estimated to be 180 mJ/K$^2$mol, which implies a significant contribution of heavy 5f electrons. 

\begin{figure}[hbt]
	\centering
		\includegraphics[width=7cm]{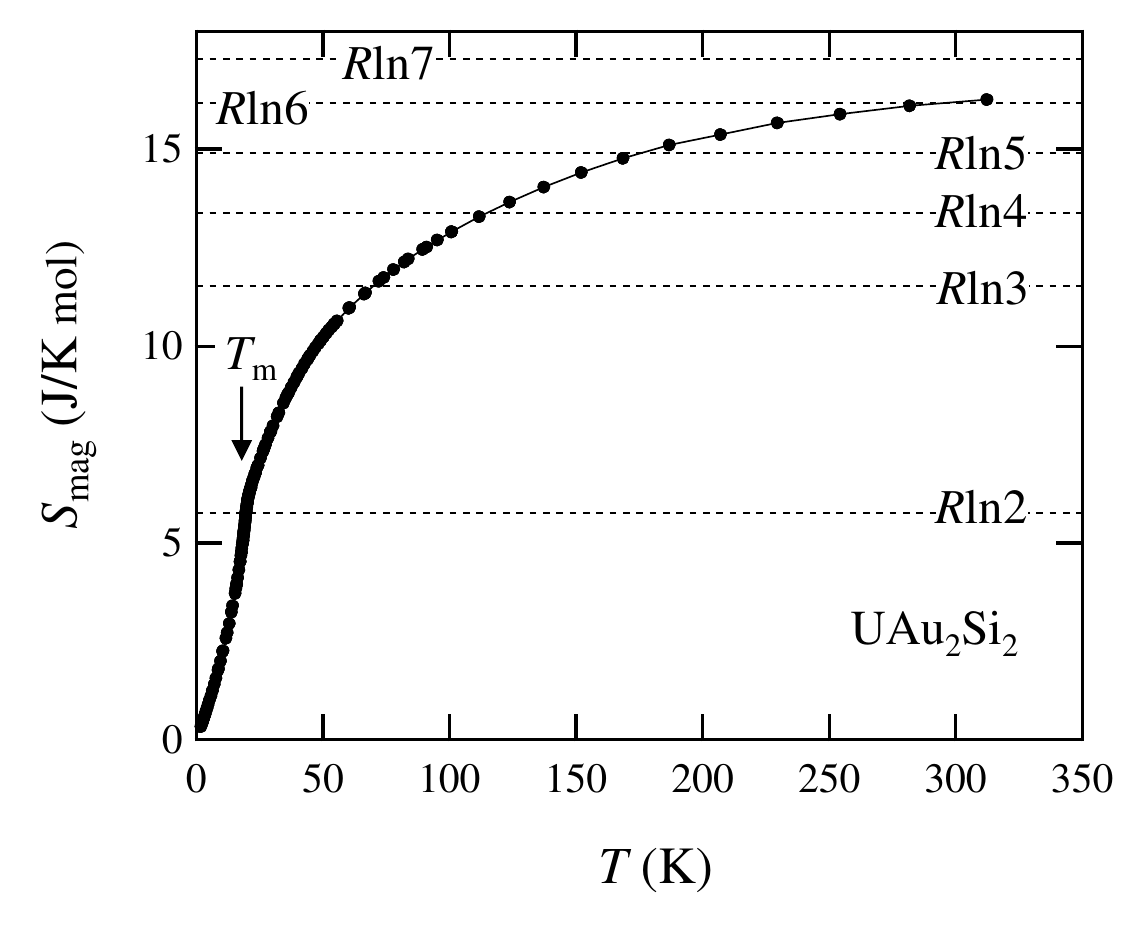}
	\caption{5f electronic contribution to the entropy of UAu$_2$Si$_2$ evaluated by integration of $C_{\rm mag}/T$, which is shown in Fig. \ref{Cmag_vs_T}. A constant value estimated by an extrapolation to $T$ = 0 is added so that the entropy goes to zero at $T$ = 0.}
	\label{Smag_vs_T}
\end{figure}

The magnetic entropy is also evaluated by integrating $C_\textrm{\scalebox{0.6}{mag}}/T$ with respect to temperature as depicted in Fig.\,\ref{Smag_vs_T}.  
It reaches ${\sim}R{\rm ln}2$ just above \tm, which suggests that one doublet or two singlets lie below the transition temperature. 
It is considered that the 5f electronic configuration of a uranium ion in a compound is (5f)$^2$ (U$^{4+}$) or (5f)$^3$ (U$^{3+}$), whose corresponding ground $J$ multiplets are $J$ = 4 or 9/2 with 9 or 10 degeneracy, respectively. In the local picture, these ground $J$ multiplets split into 5 singlets and two doublets (5f$^2$), or 5 doublets (5f$^3$) in the tetragonal crystalline electric field (CEF). 
The observed entropy release of about $R$ln6 below room temperature is thus considered to be caused by the combination of the CEF splitting, the hybridization effects between the CEF and the conduction states, and the phase transition at \tm. The isotropic feature of the effective moments deduced from the magnetization data also supports the smallness of the energy scale of the CEF splitting of this compound.

\section{\label{sec:level1}Discussions}

\subsection{Decomposition of magnetization curves}

\label{sec:discussion-50K}
\begin{figure}[hbt]
	\centering
		\includegraphics[width=7cm]{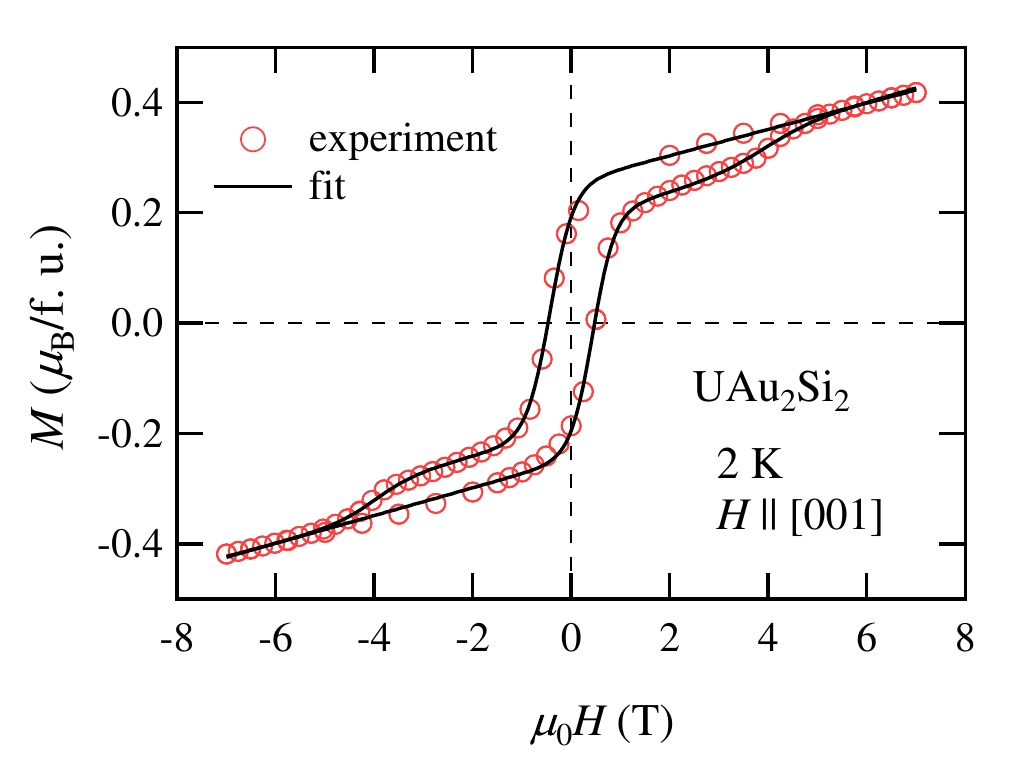}
	\caption{(Color online) Typical results of the fitting analysis of the magnetization curves of UAu$_2$Si$_2$ along the [001] axis. The fitting function is Eq. (2) which is described in the text.}
	\label{M_vs_H_fit}
\end{figure}

\begin{figure}[hbt]
	\centering
		\includegraphics[width=7cm]{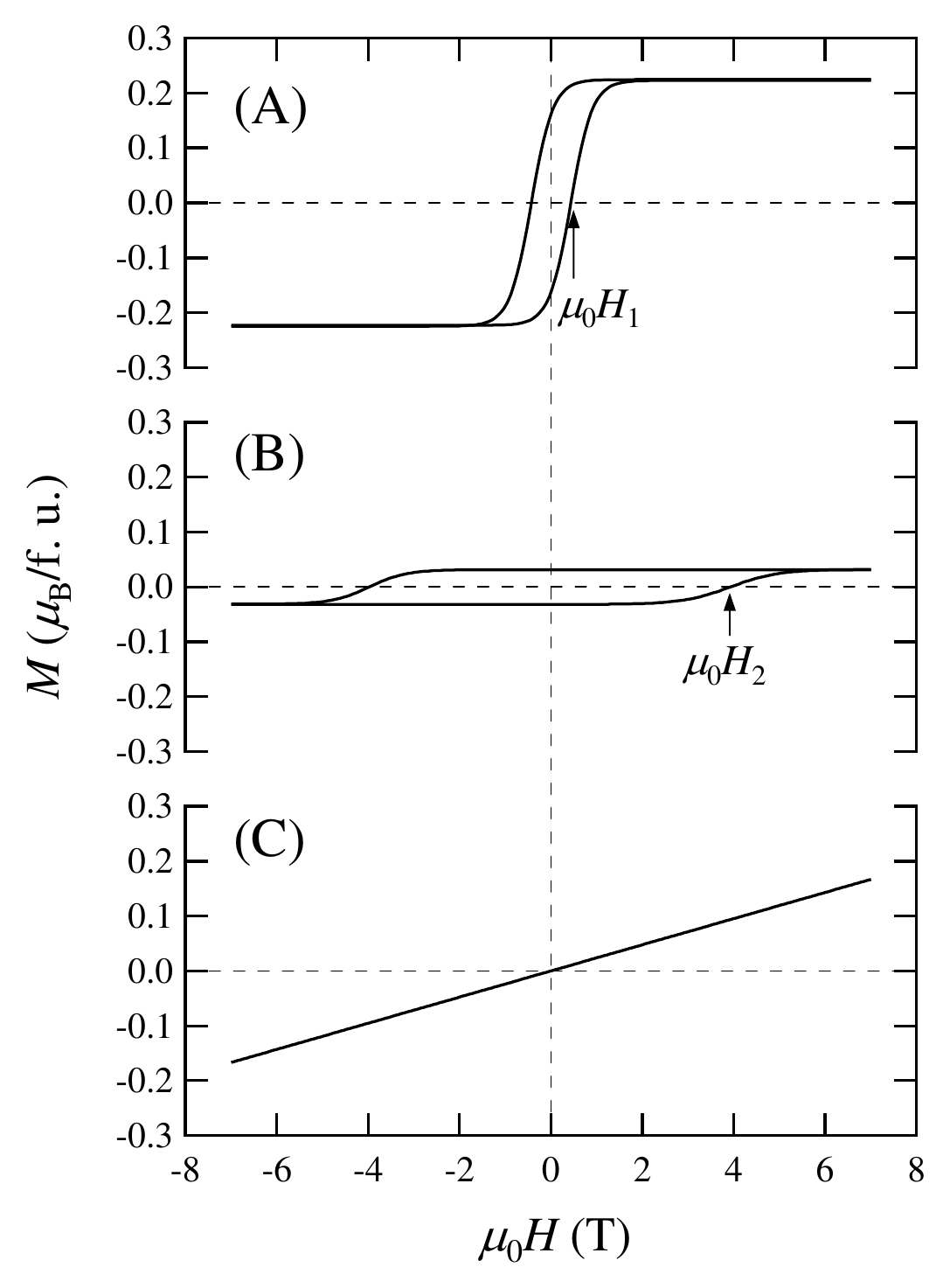}
	\caption{(A) The large FM component given by the first term of Eq. (2) using refined parameters through the curve fitting to the experimental data measured at 2 K (Fig. \ref{M_vs_H_fit}). (B) The small FM component given by the second term of Eq. (2) extracted in the same manner with (A). (C) The linear component expressed by the third term of Eq. (2), also extracted by the curve fitting. }
	\label{M_vs_H_separate}
\end{figure}

Here, we demonstrate that the nature of the weak FM component arising at $\sim$ 50 K can be understood efficiently by analyzing the $M$-$H$ curves.
As shown in Sec. \ref{sec:result-Magnetization}, the $M$-$H$ curves for fields along the [001] direction have two step-like anomalies at the fields defined as $H_1$ and $H_2$.
These complex-shaped curves can be described by summation of two different FM components. A phenomenological model based on the hyperbolic function \cite{Takacs08} successfully works in decomposition of the hysteresis loops. Namely, curve fittings using a function
\begin{eqnarray}
\label{eq_2hysteresis}
M_{\pm}(H) = &M_1&{\tanh}[\mu_{0}k_1(H{\mp}H_1)] \\ \nonumber
&+& M_2{\tanh}[\mu_{0}k_2(H{\mp}H_2)] \\ \nonumber
&+& {\mu}_{0}\chi_\textrm{\scalebox{0.6}{lin}}H + \textrm{constant} \nonumber
\end{eqnarray}
give a reasonable solution with two separated spontaneous-magnetization components. 
Here $M_{+}(H)$ and $M_{-}(H)$ represent the ascending and descending magnetization processes, respectively.
Figure \ref{M_vs_H_fit} shows that the measured magnetization is well fitted by Eq.\,(2). The first term and the second term of Eq.\,(2) are FM components with saturation magnetization of $M_1$ and $M_2$, and coercive fields of $\mu_{0}H_1$ and $\mu_{0}H_2$, respectively. The parameters $k_1$ and $k_2$ are called sheering parameters, representing the widths of magnetization switching. The third term is a linear component added from a phenomenological perspective. The fitting analysis allows us to separate the magnetization curve into three parts (except for the constant term with a small value) as displayed in Fig.\,\ref{M_vs_H_separate}.

\begin{figure}[hbt]
	\centering
		\includegraphics[width=7cm]{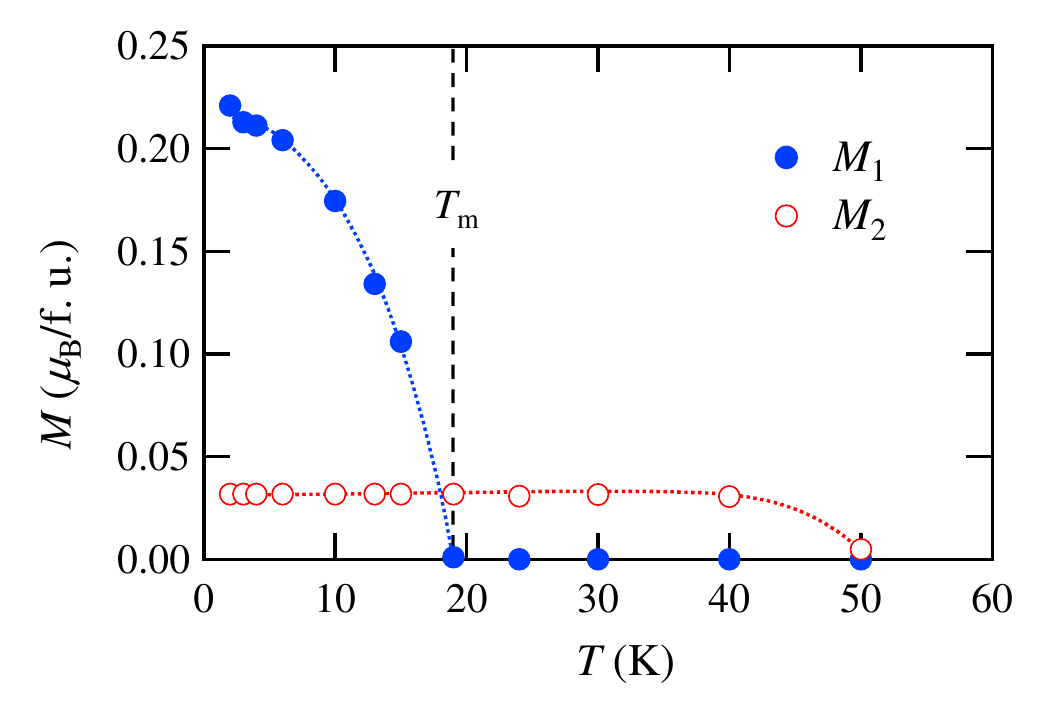}
	\caption{(Color online) Temperature dependence of the saturation magnetization of two separated FM components, $M_1$ and $M_2$, which are derived by fitting the observed magnetization loops with Eq. (2). The coercive fields, $H_1$ and $H_2$, are plotted in Fig. \ref{H1H2_vs_T}. The dotted lines are guide to the eye.}
	\label{parameters_vs_T}
\end{figure}

The temperature dependence of saturation moments $M_1$ and $M_2$, and coercive fields $H_1$ and $H_2$, are shown in Fig. \ref{parameters_vs_T} and Fig. \ref{H1H2_vs_T}, respectively.
$M_1$ was fixed to zero at the curve fitting of the data at temperature above \tm for better convergence, because it makes no significant change on the goodness of fit when compared with the case with no constraint. 
Below \tm, on the other hand, $M_2$ was fixed to a mean value of those obtained at 24 K, 34 K and 40 K for the same reason.
$M_1$ increases continuously from zero just below the transition temperature \tm, as is expected for an order parameter at a second-order phase transition.
The coercive field of the small FM component ${\mu}_{0}H_2$ shows a dramatic increase as the temperature is lowered. 
Since its magnetization $M_2$ is already saturated around 40 K, this rapid increase is considered to come from the enhancement of magnetocrystalline anisotropy.
A striking feature would be that ${\mu}_{0}H_2$ reaches approximately 4 T at 2 K, which is extremely large if compared with the behavior of typical ferromagnets. For example, it is comparable to $\sim$ 4.3 T (at 4.2 K) reported for commercial permanent magnet Co$_5$Sm \cite{Kutterer77}.
The largest coercive field among hard magnets ever known is 5.2 T at 6 K of a metal-radical polymer Co(hfac)$_{2} \cdot$BPNN \cite{Ishii08}, to the best of our knowledge.
This indicates that a considerably large magnetic anisotropy is generated in the small FM ordered state, for some reason.
We should also emphasize that $H_2$ varies smoothly and $M_2$ does not change  near \tm, indicating that the small FM component is unaffected by the phase transition occurring at \tm.

The simplest explanation of the origin of this small 50-K FM component is some extrinsic FM impurities or a second phase with different chemical component from UAu$_2$Si$_2$, because no anomaly was observed around 50 K in the specific heat.
This speculation seems to be consistent with the fact that the magnitude of the FM components shows a large sample dependence when compared with the previous reports\cite{Rebelsky91, Lin97}.
As described above, however, no detectable sign of impurities or second phases has been observed in the present XRD or EDX measurements. 
Furthermore, it is quite unlikely that a typical ferromagnetic impurity has such a huge coercive field and semi-uniaxial anisotropy in which most of the FM component aligned along the [001] direction. 
Considering all these clues together, the emergence of the 50$\,$K-FM component seems to be an independent phenomena from bulk properties of UAu$_2$Si$_2$, but not just impurities or second phases. Hence we can make one hypothesis that it originates from a portion of uranium ions whose circumstances are somehow different from those of the majority of uranium ions which cause the order at $T_{\rm m}$.
This invokes the case of URu$_2$Si$_2$, where an AFM transition occurs by applying uniaxial stress \cite{Yokoyama05}. This AFM phase is known to exist at ambient pressure with a small volume fraction, as a metastable state under the majority phase of this system, called hidden order. It is considered that such a competition of magnetic ordering can be driven by a tiny change in the lattice parameters, which is hardly detected by usual X-ray diffraction techniques.
It is also known that some crystals of URu$_2$Si$_2$ exhibit unusual FM behavior with three different onset temperatures \cite{Amitsuka07}.
Another example in the uranium 1-2-2 system would be UNi$_2$Ge$_2$, whose magnetization shows FM anomalies with a large uniaxial anisotropy, without any anomalies in other bulk properties \cite{Ning92}.
In order to clarify the origin of the 50 K-FM component, further investigation of sample dependence and measurements of other physical properties, particularly microscopic techniques such as neutron diffraction, $\mu$SR, and NMR are necessary.

\subsection{Possible type of magnetic order for $T \leq T_{\rm m}$}

As we have presented above, the magnetic moments in UAu$_2$Si$_2$ are likely to order antiferromagnetically, rather than ferromagnetically below \tm. Then what is the origin of the FM component observed along the [001] axis in the ordered state? 
We consider a case that an uncompensated AFM (UAFM) order is realized. 
The \textit{M-H} curve along the [001] direction in the ordered state can be regarded as the sum of the FM component with saturated magnetization and a component which is increasing linearly with the magnetic field expressed by the parameters $M_1$ and $\chi_\textrm{\scalebox{0.6}{lin}}$, respectively, as demonstrated in Sec. \ref{sec:discussion-50K}. This FM component can be explained by uncompensated magnetic moments along the [001] axis, based on the UAFM order model of localized U magnetic moments.

\begin{figure}[hbt]
	\centering
		\includegraphics[width=7cm]{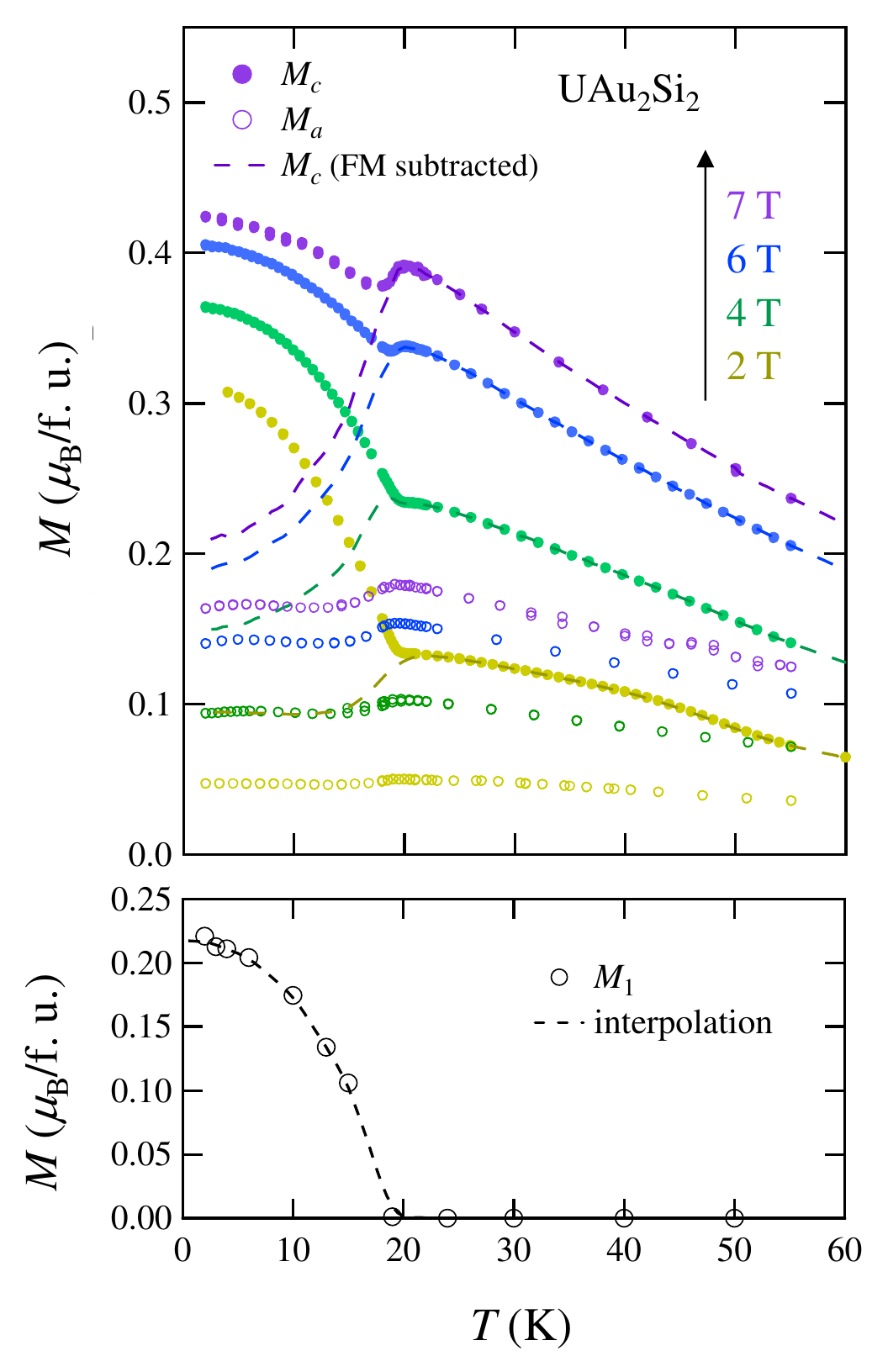}
	\caption{(Color online) (Upper panel) The temperature dependence of magnetization of UAu$_2$Si$_2$ in magnetic fields of 2 T, 4 T, 6 T, and 7 T, applied along the [001], $M_c$, (closed circles) and [100], $M_a$, (open circles) axes. The dashed lines are remaining magnetization after subtraction of the assumed FM component, which is shown in the lower panel. (Lower panel) The temperature dependence of the FM component associated with the phase transition at $T_{\rm m}$, assumed to follow that of $M_1$, which is obtained by the curve fitting using Eq. (\ref{eq_2hysteresis}).}
	\label{M_vs_T_field_subtract}
\end{figure}

This UAFM order with a ferromagnetic component also gives a rough sketch of the temperature dependence of magnetization along [001], $M_{c}(T)$. The cusp anomaly, which emerges in higher fields above 5 T, can be well accounted for by staggered components of the AFM configuration of magnetic moments. The absence of the cusp in the lower magnetic fields may be because the upturn of the FM component conceals the subtle cusp. 
To see this we assumed the upturn below \tm behaves similarly to the temperature dependence of $M_1$ as depicted in Fig. \ref{M_vs_T_field_subtract}, and then subtracted it from $M_{c}(T)$. The subtracted $M_{c}(T)$, displayed by dashed lines in the upper panel of Fig. \ref{M_vs_T_field_subtract}, shows a much larger suppression below \tm than that of $M_{a}(T)$ displayed by open circles. This indicates that there are larger staggered components of magnetic moments along the [001] axis rather than along the [100] axis. Hence this analysis suggests the UAFM order whose ordered magnetic moments are likely to be along the [001] axis.

Although this simple local-moment model successfully explains several key features of the magnetization, it is still insufficient for the actual system, because it cannot account for the component increasing in the magnetic field. We can consider that it reflects magnetic-field induced changes of a possible noncollinear magnetic structure leading to increasing projection of the U magnetic moment on the [001] axis. The noncollinear magnetic structure is also conceivable with the observed increase of the $J$\,$\parallel$\,[100] resistivity below \tm, which indicates opening a gap on the Fermi surface due to AF components developing within the basal plane. For example, one may consider a cone/umbrella or spiral magnetic structure having the [001] symmetry axis, with which the observed field-induced increase of the [001] magnetization may be accounted for by the gradual closing of the moments angle towards [001]. In addition to it, we can also consider the possibility of paramagnetic components due to the Van-Vleck paramagnetism and/or the enhanced Pauli paramagnetism. Particularly the $H_{\rm \scalebox{0.5}{m}}$-anomaly in high magnetic fields characterized by the gradual change of slope indicates existence of noncollinear component in the magnetic structure. It may be also supported by the moderate anisotropy of the effective moments. Consequently the ground state of UAu$_2$Si$_2$ is more likely to be complex and/or noncollinear AFM order, not FM order like previously reported. 

UAFM orderings with FM components have been found in several U$T_2$Si$_2$ relatives. UNi$_2$Si$_2$ is considered to show such an UAFM order with $q = (0, 0, 2/3)$ below 53 K in zero field \cite{Lin91}. UPd$_2$Si$_2$ is known to order in the same magnetic structure in a phase which appears at temperature below $\sim$ 120 K and in magnetic field above $\sim$ 0.7 T \cite{Honma98}. U(Ru$_{0.96}$Rh$_{0.04}$)$_2$Si$_2$ also has a phase in magnetic field where an UAFM ordering is realized. The recent neutron diffraction experiment in high magnetic field revealed that the propagating vector is \textit{q} = (2/3, 0, 0) \cite{Kuwahara13}.  All of these orders are collinear orders where the magnetic moments are parallel to [001]. It would be interesting to compare the low-temperature electronic state of UAu$_2$Si$_2$ with these compounds.

Here we have assumed the local moment model, but of course we can consider the possibility of an order of itinerant electrons, such as the spin-density wave. In fact, the upturn anomaly at \tm in the resistivity resembles the spin-density-wave (SDW) transition, the so-called ``Cr-like behavior'' with a commonly accepted scenario that an additional gap open on the Fermi surface caused by SDW forming. This kind of behavior of resistivity was observed also for UNi$_2$Ge$_2$ \cite{Ning92}, UCo$_2$Si$_2$ \cite{Mihalik07}, and URu$_2$Si$_2$ \cite{Palstra85, Maple86, Schlabitz86}. Future calculations of the band structure are necessary to clarify this point.

\subsection{5f-electronic properties}
\label{discussion_5f}

The experimental facts we have presented above indicate that the 5f-electronic properties which govern the magnetocrystalline anisotropy in UAu$_2$Si$_2$ significantly differ from those of the other U$T_2$Si$_2$ compounds. 
We observed moderate anisotropy in the paramagnetic state (${\it {\Theta}}_a - {\it {\Theta}}_c$ = 49 K) and the magnetic entropy that reaches ${\sim}R$ln6 at room temperature.
These features are in remarkable contrast to those of the other U$T_2$Si$_2$, which mostly exhibit stronger uniaxial magnetic anisotropy.
URu$_2$Si$_2$ and UPd$_2$Si$_2$ are such the typical examples; their magnetic susceptibility along the [001] axis clearly shows the Curie-Weiss behavior in the temperature range near the room temperature, whereas only weak temperature variations are observed in fields along [100] \cite{Palstra85,Honma93}. 
Correspondingly, their magnetic entropy at room temperature is relatively small. It is estimated to be approximately $R{\ln}3$ for these two compounds\cite{Honma93}.
The other U$T_2$Si$_2$ compounds exhibiting magnetic ordering also show similar very strong uniaxial magnetocrystalline anisotropy in paramagnetic state (except for UCu$_2$Si$_2$).

There are two principal microscopic mechanisms of magnetocrystalline anisotropy. The first one is the crystalline electric field (CEF) interaction, which is the single-ion mechanism born in the electrostatic interaction between the anisotropic crystalline electric field (potential created at the magnetic ion site by the electric charge distribution in the rest of the crystal) and the aspherical charge cloud of the magnetic electrons. The single-ion anisotropy is most often encountered in compounds based on rare-earth elements, which have well-localized 4f-electron states \cite{Herbst91}.

The second one is the hybridization effect between magnetic electrons and surrounding ligand electrons. The 5f wave functions of uranium are considerably extended in space. Consequently the 5f-electron states hybridize with ligand valence-electron states (5f-ligand hybridization \cite{Koelling85}). The strong interaction of the U 5f orbitals with surrounding ligands in the crystal caused by their large space extension implies an essentially different mechanism of magnetocrystalline anisotropy based on a two ion (U-U) interaction. A relatively simple model which leads to qualitatively realistic results has been worked out by Cooper and co-workers \cite{Cooper87} on the basis of Coqblin-Schrieffer approach to the mixing of ionic f-states and conduction-electron states \cite{Coqblin69}. The theory has been further extended so that each partially delocalized f-electron ion is coupled by anisotropic two-ion interaction giving anisotropic magnetic ordering \cite{Hu93}.

One important factor which directly correlates with the 5f-ligand hybridization is the distance between the f-ions and the non-f ligands, which can be tuned by the lattice parameters: the \textit{a} and \textit{c} parameters in the case of the tetragonal structure. 
The smaller distance makes the hybridization stronger.
Another factor is the number of d-electrons of the transition-metal ions.
It is commonly believed that the increase of the d-band filling weakens the d-f hybridization, because the energy of d-band is considered to be pulled down away from the Fermi level by increasing the filling, resulting in a smaller overlap of the d and f bands \cite{Endstra93,Sandratskii94}. 
Since UAu$_2$Si$_2$ has the largest lattice parameter of $a$ ${\sim}$ 10.3 {\AA} \ and the largest d-band filling in the 5d systems of U$T_2$Si$_2$, the weakest 5f-d hybridization within the U$T_2$Si$_2$ compounds can be expected.
This consideration is consistent with the present results just mentioned above.
Note that UCu$_2$Si$_2$, which is characterized by the highest d-band filling of the 3d transition metals leading to the minimized 5f-3d hybridization \cite{Sandratskii94} and exhibits large magnetic entropy at room temperature of ${\sim}R$ln9 \cite{Matsuda05}, represents a somewhat intermediate case (${\it {\Theta}}_a-{\it {\Theta}}_c$ $\sim$ 180 K) between the moderate anisotropy in UAu$_2$Si$_2$ and the very strong anisotropy of other U$T_2$Si$_2$ compounds having a magnetically ordered ground state.
The fact that UAu$_2$Si$_2$ and UCu$_2$Si$_2$ exhibit moderate anisotropy is considered to be due to a dramatically reduced energy scale of the two-ion interaction reflecting the minimized 5f-d hybridization as a consequence of their large lattice parameters and characteristics of high d-band filling.

A local character of the 5f electrons in UAu$_2$Si$_2$ can be deduced from the magnetic susceptibility, which follows the Curie-Weiss's law at high temperatures above $\sim$ 60 K. 
Moreover, the electrical resistivity shows roughly the -ln$T$ behavior from room temperature down to ${\sim}$ 60 K.
If this is due to the Kondo effect, this also means existence of local 5f electron magnetic moments.
On the other hand, at lower temperature the 5f electrons behave like itinerant electrons, as seen in the large ${\gamma}$ and $A$ values, with which we can put UAu$_2$Si$_2$ on the Kadowaki-Woods plot for the typical heavy fermion compounds, $A/{\gamma}^2 = 1 \times 10^{-5} {\mu}{\Omega}{\rm cm}({\rm K} {\rm mol}/{\rm m} {\rm J})^2$.
Although these values should be interpreted carefully, because they are deduced from the data from the magnetically ordered state, these experimental facts strongly suggest that the low-temperature state of this compound is described by the Fermi-liquid theory with heavy 5f itinerant electrons.

\section{Conclusions}

We have succeeded in growing a single-crystalline sample of a uranium intermetallic compound UAu$_2$Si$_2$, and made a characterization of it through measurements of specific heat, electrical resistivity and magnetization. 
We have confirmed that a second-order phase transition occurs at 19 K, with a spontaneous magnetization only along [001] direction. This order is considered to be spin-uncompensated and/or noncollinear AFM order, not FM order as  believed in the previous reports.
The magnetic field-temperature phase diagram with the applied field along the [001] direction suggests an existence of another magnetically ordered phase in magnetic fields above $\sim$ 8 T; UAu$_2$Si$_2$ might have multiple magnetic phases with different magnetic structures like some of other 1-2-2 relatives such as UPd$_2$Si$_2$ and URu$_2$Si$_2$.
The weak one-ion magnetic anisotropy in the paramagnetic range and the large magnetic entropy at room temperature have been revealed. These experimental facts suggest that the relatively low energy of the 5f-ligand hybridization as well as the CEF effects induces the interaction responsible for the magnetic anisotropy in UAu$_2$Si$_2$. This is a vivid contrast to the cases of most other U$T_2$Si$_2$ compounds.

The origin of weak FM component arising at about 50 K is still an open issue. Its peculiar properties, such as semi-uniaxial anisotropy along the crystalline [001]-axis and the giant coercive field, imply that it is not just an impurity or second phase contribution. Meanwhile, the absence of the specific-heat anomaly and separability of the magnetization curve suggests that it is independent phenomena from the bulk phase transition at 19 K. This incompatibility should be resolved through more detailed experiments in future. 
The improvement of the quality of the single-crystalline samples and microscopic experiments to determine the magnetic structure will shed light on the nature of the peculiar magnetism of UAu$_2$Si$_2$.

\begin{acknowledgments}
We are grateful to Prof.${\:}{\:}$Mohsen Abd Elmeguid for fruitful discussions. The present research was supported by JSPS Grants-in-Aid for Scientific Research (KAKENHI) Grant No. 15H05882, 15H05885 and 15K21732 (J-Physics), and for the Strategic Young Researcher Overseas Visits Program for Accelerating Brain Circulation. The Prague group was supported by the Czech Science Foundation by the Grant No. P204/15/03777S. Experiments with single crystals were performed in MLTL (http://mltl.eu/) which is supported within the program of Czech Research Infrastructures (Project No. LM2011025).
\end{acknowledgments}


\begin{thebibliography}{46}%
\makeatletter
\providecommand \@ifxundefined [1]{%
 \@ifx{#1\undefined}
}%
\providecommand \@ifnum [1]{%
 \ifnum #1\expandafter \@firstoftwo
 \else \expandafter \@secondoftwo
 \fi
}%
\providecommand \@ifx [1]{%
 \ifx #1\expandafter \@firstoftwo
 \else \expandafter \@secondoftwo
 \fi
}%
\providecommand \natexlab [1]{#1}%
\providecommand \enquote  [1]{``#1''}%
\providecommand \bibnamefont  [1]{#1}%
\providecommand \bibfnamefont [1]{#1}%
\providecommand \citenamefont [1]{#1}%
\providecommand \href@noop [0]{\@secondoftwo}%
\providecommand \href [0]{\begingroup \@sanitize@url \@href}%
\providecommand \@href[1]{\@@startlink{#1}\@@href}%
\providecommand \@@href[1]{\endgroup#1\@@endlink}%
\providecommand \@sanitize@url [0]{\catcode `\\12\catcode `\$12\catcode
  `\&12\catcode `\#12\catcode `\^12\catcode `\_12\catcode `\%12\relax}%
\providecommand \@@startlink[1]{}%
\providecommand \@@endlink[0]{}%
\providecommand \url  [0]{\begingroup\@sanitize@url \@url }%
\providecommand \@url [1]{\endgroup\@href {#1}{\urlprefix }}%
\providecommand \urlprefix  [0]{URL }%
\providecommand \Eprint [0]{\href }%
\providecommand \doibase [0]{http://dx.doi.org/}%
\providecommand \selectlanguage [0]{\@gobble}%
\providecommand \bibinfo  [0]{\@secondoftwo}%
\providecommand \bibfield  [0]{\@secondoftwo}%
\providecommand \translation [1]{[#1]}%
\providecommand \BibitemOpen [0]{}%
\providecommand \bibitemStop [0]{}%
\providecommand \bibitemNoStop [0]{.\EOS\space}%
\providecommand \EOS [0]{\spacefactor3000\relax}%
\providecommand \BibitemShut  [1]{\csname bibitem#1\endcsname}%
\let\auto@bib@innerbib\@empty
\bibitem [{\citenamefont {{Matsuka}}\ \emph {et~al.}(2003)\citenamefont
  {{Matsuka}}, \citenamefont {{Metoki}}, \citenamefont {{Haga}}, \citenamefont
  {{Ikeda}}, \citenamefont {{Okubo}}, \citenamefont {{Sugiyama}}, \citenamefont
  {{Nakamura}}, \citenamefont {{Kindo}}, \citenamefont {{Kaneko}},
  \citenamefont {{Nakamura}}, \citenamefont {{Yamamoto}},\ and\ \citenamefont
  {{{\=O}nuki}}}]{Matsuda03}%
  \BibitemOpen
  \bibfield  {author} {\bibinfo {author} {\bibfnamefont {T.~D.}\ \bibnamefont
  {{Matsuka}}}, \bibinfo {author} {\bibfnamefont {N.}~\bibnamefont {{Metoki}}},
  \bibinfo {author} {\bibfnamefont {Y.}~\bibnamefont {{Haga}}}, \bibinfo
  {author} {\bibfnamefont {S.}~\bibnamefont {{Ikeda}}}, \bibinfo {author}
  {\bibfnamefont {T.}~\bibnamefont {{Okubo}}}, \bibinfo {author} {\bibfnamefont
  {K.}~\bibnamefont {{Sugiyama}}}, \bibinfo {author} {\bibfnamefont
  {N.}~\bibnamefont {{Nakamura}}}, \bibinfo {author} {\bibfnamefont
  {K.}~\bibnamefont {{Kindo}}}, \bibinfo {author} {\bibfnamefont
  {K.}~\bibnamefont {{Kaneko}}}, \bibinfo {author} {\bibfnamefont
  {A.}~\bibnamefont {{Nakamura}}}, \bibinfo {author} {\bibfnamefont
  {E.}~\bibnamefont {{Yamamoto}}}, \ and\ \bibinfo {author} {\bibfnamefont
  {Y.}~\bibnamefont {{{\=O}nuki}}},\ }\href@noop {} {\bibfield  {journal}
  {\bibinfo  {journal} {J. Phys. Soc. Jpn.}\ }\textbf {\bibinfo {volume}
  {72}},\ \bibinfo {pages} {122} (\bibinfo {year} {2003})}\BibitemShut
  {NoStop}%
\bibitem [{\citenamefont {{Szytula}}\ \emph {et~al.}(1988)\citenamefont
  {{Szytula}}, \citenamefont {{Siek}}, \citenamefont {{Leciefewicz}},
  \citenamefont {{Zygmunt}},\ and\ \citenamefont {{Ban}}}]{Szytula88}%
  \BibitemOpen
  \bibfield  {author} {\bibinfo {author} {\bibfnamefont {A.}~\bibnamefont
  {{Szytula}}}, \bibinfo {author} {\bibfnamefont {S.}~\bibnamefont {{Siek}}},
  \bibinfo {author} {\bibfnamefont {J.}~\bibnamefont {{Leciefewicz}}}, \bibinfo
  {author} {\bibfnamefont {A.}~\bibnamefont {{Zygmunt}}}, \ and\ \bibinfo
  {author} {\bibfnamefont {Z.}~\bibnamefont {{Ban}}},\ }\href@noop {}
  {\bibfield  {journal} {\bibinfo  {journal} {J. Phys. Chem. Solids}\ }\textbf
  {\bibinfo {volume} {49}},\ \bibinfo {pages} {1113} (\bibinfo {year}
  {1988})}\BibitemShut {NoStop}%
\bibitem [{\citenamefont {{Chelmicki}}\ \emph {et~al.}(1985)\citenamefont
  {{Chelmicki}}, \citenamefont {{Leciefewicz}},\ and\ \citenamefont
  {{Zygmunt}}}]{Chelmicki85}%
  \BibitemOpen
  \bibfield  {author} {\bibinfo {author} {\bibfnamefont {L.}~\bibnamefont
  {{Chelmicki}}}, \bibinfo {author} {\bibfnamefont {J.}~\bibnamefont
  {{Leciefewicz}}}, \ and\ \bibinfo {author} {\bibfnamefont {A.}~\bibnamefont
  {{Zygmunt}}},\ }\href@noop {} {\bibfield  {journal} {\bibinfo  {journal} {J.
  Phys. Chem. Solids}\ }\textbf {\bibinfo {volume} {46}},\ \bibinfo {pages}
  {579} (\bibinfo {year} {1985})}\BibitemShut {NoStop}%
\bibitem [{\citenamefont {{Lin}}\ \emph {et~al.}(1991)\citenamefont {{Lin}},
  \citenamefont {{Rebelsky}}, \citenamefont {{Collins}}, \citenamefont
  {{Garrett}},\ and\ \citenamefont {{Buyers}}}]{Lin91}%
  \BibitemOpen
  \bibfield  {author} {\bibinfo {author} {\bibfnamefont {H.}~\bibnamefont
  {{Lin}}}, \bibinfo {author} {\bibfnamefont {L.}~\bibnamefont {{Rebelsky}}},
  \bibinfo {author} {\bibfnamefont {M.~F.}\ \bibnamefont {{Collins}}}, \bibinfo
  {author} {\bibfnamefont {J.~D.}\ \bibnamefont {{Garrett}}}, \ and\ \bibinfo
  {author} {\bibfnamefont {W.~J.~L.}\ \bibnamefont {{Buyers}}},\ }\href@noop {}
  {\bibfield  {journal} {\bibinfo  {journal} {Phys. Rev. B}\ }\textbf {\bibinfo
  {volume} {43}},\ \bibinfo {pages} {13232} (\bibinfo {year}
  {1991})}\BibitemShut {NoStop}%
\bibitem [{\citenamefont {{Matsuda}}\ \emph {et~al.}(2005)\citenamefont
  {{Matsuda}}, \citenamefont {{Haga}}, \citenamefont {{Ikeda}}, \citenamefont
  {{Galatanu}}, \citenamefont {{Yamamoto}}, \citenamefont {{Shishido}},
  \citenamefont {{Yamada}}, \citenamefont {{Yamamura}}, \citenamefont {{Hedo}},
  \citenamefont {{Uwatoko}}, \citenamefont {{Matsumoto}}, \citenamefont
  {{Tada}}, \citenamefont {{Noguchi}}, \citenamefont {{Sugimoto}},
  \citenamefont {{Kuwahara}}, \citenamefont {{Iwasa}}, \citenamefont {{Kohgi}},
  \citenamefont {{Settai}},\ and\ \citenamefont {{{\=O}nuki}}}]{Matsuda05}%
  \BibitemOpen
  \bibfield  {author} {\bibinfo {author} {\bibfnamefont {T.~D.}\ \bibnamefont
  {{Matsuda}}}, \bibinfo {author} {\bibfnamefont {Y.}~\bibnamefont {{Haga}}},
  \bibinfo {author} {\bibfnamefont {S.}~\bibnamefont {{Ikeda}}}, \bibinfo
  {author} {\bibfnamefont {A.}~\bibnamefont {{Galatanu}}}, \bibinfo {author}
  {\bibfnamefont {E.}~\bibnamefont {{Yamamoto}}}, \bibinfo {author}
  {\bibfnamefont {H.}~\bibnamefont {{Shishido}}}, \bibinfo {author}
  {\bibfnamefont {M.}~\bibnamefont {{Yamada}}}, \bibinfo {author}
  {\bibfnamefont {J.}~\bibnamefont {{Yamamura}}}, \bibinfo {author}
  {\bibfnamefont {M.}~\bibnamefont {{Hedo}}}, \bibinfo {author} {\bibfnamefont
  {Y.}~\bibnamefont {{Uwatoko}}}, \bibinfo {author} {\bibfnamefont
  {T.}~\bibnamefont {{Matsumoto}}}, \bibinfo {author} {\bibfnamefont
  {T.}~\bibnamefont {{Tada}}}, \bibinfo {author} {\bibfnamefont
  {S.}~\bibnamefont {{Noguchi}}}, \bibinfo {author} {\bibfnamefont
  {T.}~\bibnamefont {{Sugimoto}}}, \bibinfo {author} {\bibfnamefont
  {K.}~\bibnamefont {{Kuwahara}}}, \bibinfo {author} {\bibfnamefont
  {K.}~\bibnamefont {{Iwasa}}}, \bibinfo {author} {\bibfnamefont
  {M.}~\bibnamefont {{Kohgi}}}, \bibinfo {author} {\bibfnamefont
  {R.}~\bibnamefont {{Settai}}}, \ and\ \bibinfo {author} {\bibfnamefont
  {Y.}~\bibnamefont {{{\=O}nuki}}},\ }\href@noop {} {\bibfield  {journal}
  {\bibinfo  {journal} {J. Phys. Soc. Jpn.}\ }\textbf {\bibinfo {volume}
  {74}},\ \bibinfo {pages} {1552} (\bibinfo {year} {2005})}\BibitemShut
  {NoStop}%
\bibitem [{\citenamefont {{Honda}}\ \emph {et~al.}(2006)\citenamefont
  {{Honda}}, \citenamefont {{Metoki}}, \citenamefont {{Matsuda}}, \citenamefont
  {{Haga}},\ and\ \citenamefont {{{\=O}nuki}}}]{Honda06}%
  \BibitemOpen
  \bibfield  {author} {\bibinfo {author} {\bibfnamefont {F.}~\bibnamefont
  {{Honda}}}, \bibinfo {author} {\bibfnamefont {N.}~\bibnamefont {{Metoki}}},
  \bibinfo {author} {\bibfnamefont {T.~D.}\ \bibnamefont {{Matsuda}}}, \bibinfo
  {author} {\bibfnamefont {Y.}~\bibnamefont {{Haga}}}, \ and\ \bibinfo {author}
  {\bibfnamefont {Y.}~\bibnamefont {{{\=O}nuki}}},\ }\href@noop {} {\bibfield
  {journal} {\bibinfo  {journal} {J. Phys. Cond. Matt.}\ }\textbf {\bibinfo
  {volume} {18}},\ \bibinfo {pages} {479} (\bibinfo {year} {2006})}\BibitemShut
  {NoStop}%
\bibitem [{\citenamefont {{Ptasiewicz-Bak}}\ \emph {et~al.}(1981)\citenamefont
  {{Ptasiewicz-Bak}}, \citenamefont {{Leciejewicz}},\ and\ \citenamefont
  {{Zygmunt}}}]{Bak81}%
  \BibitemOpen
  \bibfield  {author} {\bibinfo {author} {\bibfnamefont {H.}~\bibnamefont
  {{Ptasiewicz-Bak}}}, \bibinfo {author} {\bibfnamefont {J.}~\bibnamefont
  {{Leciejewicz}}}, \ and\ \bibinfo {author} {\bibfnamefont {A.}~\bibnamefont
  {{Zygmunt}}},\ }\href@noop {} {\bibfield  {journal} {\bibinfo  {journal} {J.
  Phys. F: Metal Phys.}\ }\textbf {\bibinfo {volume} {11}},\ \bibinfo {pages}
  {225} (\bibinfo {year} {1981})}\BibitemShut {NoStop}%
\bibitem [{\citenamefont {{Palstra}}\ \emph {et~al.}(1986)\citenamefont
  {{Palstra}}, \citenamefont {{Menovsky}}, \citenamefont {{Nieruwenhuys}},\
  and\ \citenamefont {{Mydosh}}}]{Palstra86}%
  \BibitemOpen
  \bibfield  {author} {\bibinfo {author} {\bibfnamefont {T.~T.~M.}\
  \bibnamefont {{Palstra}}}, \bibinfo {author} {\bibfnamefont {A.~A.}\
  \bibnamefont {{Menovsky}}}, \bibinfo {author} {\bibfnamefont {G.~J.}\
  \bibnamefont {{Nieruwenhuys}}}, \ and\ \bibinfo {author} {\bibfnamefont
  {J.~A.}\ \bibnamefont {{Mydosh}}},\ }\href@noop {} {\bibfield  {journal}
  {\bibinfo  {journal} {J. Magn. Magn. Mater.}\ }\textbf {\bibinfo {volume}
  {54-57}},\ \bibinfo {pages} {435} (\bibinfo {year} {1986})}\BibitemShut
  {NoStop}%
\bibitem [{\citenamefont {{Honma}}\ \emph {et~al.}(1998)\citenamefont
  {{Honma}}, \citenamefont {{Amitsuka}}, \citenamefont {{Yasunami}},
  \citenamefont {{Tenya}}, \citenamefont {{Sakakibara}}, \citenamefont
  {{Mitamura}}, \citenamefont {{Goto}}, \citenamefont {{Kido}}, \citenamefont
  {{Miyako}}, \citenamefont {{Sugiyama}},\ and\ \citenamefont
  {{Date}}}]{Honma98}%
  \BibitemOpen
  \bibfield  {author} {\bibinfo {author} {\bibfnamefont {T.}~\bibnamefont
  {{Honma}}}, \bibinfo {author} {\bibfnamefont {H.}~\bibnamefont {{Amitsuka}}},
  \bibinfo {author} {\bibfnamefont {S.}~\bibnamefont {{Yasunami}}}, \bibinfo
  {author} {\bibfnamefont {K.}~\bibnamefont {{Tenya}}}, \bibinfo {author}
  {\bibfnamefont {T.}~\bibnamefont {{Sakakibara}}}, \bibinfo {author}
  {\bibfnamefont {H.}~\bibnamefont {{Mitamura}}}, \bibinfo {author}
  {\bibfnamefont {T.}~\bibnamefont {{Goto}}}, \bibinfo {author} {\bibfnamefont
  {S.}~\bibnamefont {{Kido}}, \bibfnamefont {G.and~{Kawarazaki}}}, \bibinfo
  {author} {\bibfnamefont {Y.}~\bibnamefont {{Miyako}}}, \bibinfo {author}
  {\bibfnamefont {K.}~\bibnamefont {{Sugiyama}}}, \ and\ \bibinfo {author}
  {\bibfnamefont {M.}~\bibnamefont {{Date}}},\ }\href@noop {} {\bibfield
  {journal} {\bibinfo  {journal} {J. Phys. Soc. Jpn.}\ }\textbf {\bibinfo
  {volume} {67}},\ \bibinfo {pages} {1017} (\bibinfo {year}
  {1998})}\BibitemShut {NoStop}%
\bibitem [{\citenamefont {{Dirkmaat}}\ \emph {et~al.}(1990)\citenamefont
  {{Dirkmaat}}, \citenamefont {{Endstra}}, \citenamefont {{Knetsch}},
  \citenamefont {{Menovsky}}, \citenamefont {{Nieuwenhuys}},\ and\
  \citenamefont {{Mydosh}}}]{Dirkmaat90}%
  \BibitemOpen
  \bibfield  {author} {\bibinfo {author} {\bibfnamefont {A.~J.}\ \bibnamefont
  {{Dirkmaat}}}, \bibinfo {author} {\bibfnamefont {T.}~\bibnamefont
  {{Endstra}}}, \bibinfo {author} {\bibfnamefont {E.~A.}\ \bibnamefont
  {{Knetsch}}}, \bibinfo {author} {\bibfnamefont {A.~A.}\ \bibnamefont
  {{Menovsky}}}, \bibinfo {author} {\bibfnamefont {G.~J.}\ \bibnamefont
  {{Nieuwenhuys}}}, \ and\ \bibinfo {author} {\bibfnamefont {J.~A.}\
  \bibnamefont {{Mydosh}}},\ }\href@noop {} {\bibfield  {journal} {\bibinfo
  {journal} {J. Magn. Magn. Mater.}\ }\textbf {\bibinfo {volume} {84}},\
  \bibinfo {pages} {143} (\bibinfo {year} {1990})}\BibitemShut {NoStop}%
\bibitem [{\citenamefont {{Verni{\'e}re}}\ \emph {et~al.}(1996)\citenamefont
  {{Verni{\'e}re}}, \citenamefont {{Raymond}}, \citenamefont {{Boucherle}},
  \citenamefont {{Lejay}}, \citenamefont {{F\r{a}k}}, \citenamefont
  {{Flouquet}},\ and\ \citenamefont {{Mignot}}}]{Verniere96}%
  \BibitemOpen
  \bibfield  {author} {\bibinfo {author} {\bibfnamefont {A.}~\bibnamefont
  {{Verni{\'e}re}}}, \bibinfo {author} {\bibfnamefont {S.}~\bibnamefont
  {{Raymond}}}, \bibinfo {author} {\bibfnamefont {J.~X.}\ \bibnamefont
  {{Boucherle}}}, \bibinfo {author} {\bibfnamefont {P.}~\bibnamefont
  {{Lejay}}}, \bibinfo {author} {\bibfnamefont {B.}~\bibnamefont {{F\r{a}k}}},
  \bibinfo {author} {\bibfnamefont {J.}~\bibnamefont {{Flouquet}}}, \ and\
  \bibinfo {author} {\bibfnamefont {J.~M.}\ \bibnamefont {{Mignot}}},\
  }\href@noop {} {\bibfield  {journal} {\bibinfo  {journal} {J. Magn. Magn.
  Mater.}\ }\textbf {\bibinfo {volume} {153}},\ \bibinfo {pages} {55} (\bibinfo
  {year} {1996})}\BibitemShut {NoStop}%
\bibitem [{\citenamefont {{Steeman}}\ \emph {et~al.}(1990)\citenamefont
  {{Steeman}}, \citenamefont {{Frikkee}}, \citenamefont {{Mentink}},
  \citenamefont {{Menovsky}}, \citenamefont {{Nieuwenhuys}},\ and\
  \citenamefont {{Mydosh}}}]{Steeman90}%
  \BibitemOpen
  \bibfield  {author} {\bibinfo {author} {\bibfnamefont {R.~A.}\ \bibnamefont
  {{Steeman}}}, \bibinfo {author} {\bibfnamefont {E.}~\bibnamefont
  {{Frikkee}}}, \bibinfo {author} {\bibfnamefont {S.~A.~M.}\ \bibnamefont
  {{Mentink}}}, \bibinfo {author} {\bibfnamefont {A.~A.}\ \bibnamefont
  {{Menovsky}}}, \bibinfo {author} {\bibfnamefont {G.~J.}\ \bibnamefont
  {{Nieuwenhuys}}}, \ and\ \bibinfo {author} {\bibfnamefont {J.~A.}\
  \bibnamefont {{Mydosh}}},\ }\href@noop {} {\bibfield  {journal} {\bibinfo
  {journal} {J. Phys. Cond. Matt.}\ }\textbf {\bibinfo {volume} {2}},\ \bibinfo
  {pages} {4059} (\bibinfo {year} {1990})}\BibitemShut {NoStop}%
\bibitem [{\citenamefont {{S\"{u}llow}}\ \emph {et~al.}(2008)\citenamefont
  {{S\"{u}llow}}, \citenamefont {{Otop}}, \citenamefont {{Loose}},
  \citenamefont {{Klenke}}, \citenamefont {{Prokhnenko}}, \citenamefont
  {{Feyerherm}}, \citenamefont {{Hendrikx}}, \citenamefont {{Mydosh}},\ and\
  \citenamefont {{Amitsuka}}}]{Sullow08}%
  \BibitemOpen
  \bibfield  {author} {\bibinfo {author} {\bibfnamefont {S.}~\bibnamefont
  {{S\"{u}llow}}}, \bibinfo {author} {\bibfnamefont {A.}~\bibnamefont
  {{Otop}}}, \bibinfo {author} {\bibfnamefont {A.}~\bibnamefont {{Loose}}},
  \bibinfo {author} {\bibfnamefont {J.}~\bibnamefont {{Klenke}}}, \bibinfo
  {author} {\bibfnamefont {O.}~\bibnamefont {{Prokhnenko}}}, \bibinfo {author}
  {\bibfnamefont {R.}~\bibnamefont {{Feyerherm}}}, \bibinfo {author}
  {\bibfnamefont {R.~W.~A.}\ \bibnamefont {{Hendrikx}}}, \bibinfo {author}
  {\bibfnamefont {J.~A.}\ \bibnamefont {{Mydosh}}}, \ and\ \bibinfo {author}
  {\bibfnamefont {H.}~\bibnamefont {{Amitsuka}}},\ }\href@noop {} {\bibfield
  {journal} {\bibinfo  {journal} {J. Phys. Soc. Jpn.}\ }\textbf {\bibinfo
  {volume} {77}},\ \bibinfo {pages} {024708} (\bibinfo {year}
  {2008})}\BibitemShut {NoStop}%
\bibitem [{\citenamefont {{Lin}}\ \emph {et~al.}(1997)\citenamefont {{Lin}},
  \citenamefont {{Hwang}}, \citenamefont {{Wur}}, \citenamefont {{Hsu}},\ and\
  \citenamefont {{Tien}}}]{Lin97}%
  \BibitemOpen
  \bibfield  {author} {\bibinfo {author} {\bibfnamefont {K.~J.}\ \bibnamefont
  {{Lin}}}, \bibinfo {author} {\bibfnamefont {J.~S.}\ \bibnamefont {{Hwang}}},
  \bibinfo {author} {\bibfnamefont {C.~S.}\ \bibnamefont {{Wur}}}, \bibinfo
  {author} {\bibfnamefont {R.}~\bibnamefont {{Hsu}}}, \ and\ \bibinfo {author}
  {\bibfnamefont {C.}~\bibnamefont {{Tien}}},\ }\href@noop {} {\bibfield
  {journal} {\bibinfo  {journal} {Solid State Commun.}\ }\textbf {\bibinfo
  {volume} {103}},\ \bibinfo {pages} {185} (\bibinfo {year}
  {1997})}\BibitemShut {NoStop}%
\bibitem [{\citenamefont {{Palstra}}\ \emph {et~al.}(1985)\citenamefont
  {{Palstra}}, \citenamefont {{Menovsky}}, \citenamefont {{van den Berg}},
  \citenamefont {{Dirkmaat}}, \citenamefont {{Kes}}, \citenamefont
  {{Nieuwenhuys}},\ and\ \citenamefont {{Mydosh}}}]{Palstra85}%
  \BibitemOpen
  \bibfield  {author} {\bibinfo {author} {\bibfnamefont {T.~T.~M.}\
  \bibnamefont {{Palstra}}}, \bibinfo {author} {\bibfnamefont {A.~A.}\
  \bibnamefont {{Menovsky}}}, \bibinfo {author} {\bibfnamefont
  {J.}~\bibnamefont {{van den Berg}}}, \bibinfo {author} {\bibfnamefont
  {A.~J.}\ \bibnamefont {{Dirkmaat}}}, \bibinfo {author} {\bibfnamefont
  {P.~H.}\ \bibnamefont {{Kes}}}, \bibinfo {author} {\bibfnamefont {G.~J.}\
  \bibnamefont {{Nieuwenhuys}}}, \ and\ \bibinfo {author} {\bibfnamefont
  {J.~A.}\ \bibnamefont {{Mydosh}}},\ }\href@noop {} {\bibfield  {journal}
  {\bibinfo  {journal} {Phys. Rev. Lett.}\ }\textbf {\bibinfo {volume} {55}},\
  \bibinfo {pages} {2727} (\bibinfo {year} {1985})}\BibitemShut {NoStop}%
\bibitem [{\citenamefont {{Maple}}\ \emph {et~al.}(1986)\citenamefont
  {{Maple}}, \citenamefont {{Chen}}, \citenamefont {{Dalichaouch}},
  \citenamefont {{Kohara}}, \citenamefont {{Rossel}}, \citenamefont
  {{Torikachvili}}, \citenamefont {{McElfresh}},\ and\ \citenamefont
  {{Thompson}}}]{Maple86}%
  \BibitemOpen
  \bibfield  {author} {\bibinfo {author} {\bibfnamefont {M.~B.}\ \bibnamefont
  {{Maple}}}, \bibinfo {author} {\bibfnamefont {J.~W.}\ \bibnamefont {{Chen}}},
  \bibinfo {author} {\bibfnamefont {Y.}~\bibnamefont {{Dalichaouch}}}, \bibinfo
  {author} {\bibfnamefont {T.}~\bibnamefont {{Kohara}}}, \bibinfo {author}
  {\bibfnamefont {C.}~\bibnamefont {{Rossel}}}, \bibinfo {author}
  {\bibfnamefont {M.~S.}\ \bibnamefont {{Torikachvili}}}, \bibinfo {author}
  {\bibfnamefont {M.~W.}\ \bibnamefont {{McElfresh}}}, \ and\ \bibinfo {author}
  {\bibfnamefont {J.~D.}\ \bibnamefont {{Thompson}}},\ }\href@noop {}
  {\bibfield  {journal} {\bibinfo  {journal} {Phys. Rev. Lett.}\ }\textbf
  {\bibinfo {volume} {56}},\ \bibinfo {pages} {185} (\bibinfo {year}
  {1986})}\BibitemShut {NoStop}%
\bibitem [{\citenamefont {{Schlabitz}}\ \emph {et~al.}(1986)\citenamefont
  {{Schlabitz}}, \citenamefont {{Baumaan}}, \citenamefont {{Pollit}},
  \citenamefont {{Rauchschwalbe}}, \citenamefont {{Mayer}}, \citenamefont
  {{Ahlheim}},\ and\ \citenamefont {{Bredl}}}]{Schlabitz86}%
  \BibitemOpen
  \bibfield  {author} {\bibinfo {author} {\bibfnamefont {W.}~\bibnamefont
  {{Schlabitz}}}, \bibinfo {author} {\bibfnamefont {J.}~\bibnamefont
  {{Baumaan}}}, \bibinfo {author} {\bibfnamefont {B.}~\bibnamefont {{Pollit}}},
  \bibinfo {author} {\bibfnamefont {U.}~\bibnamefont {{Rauchschwalbe}}},
  \bibinfo {author} {\bibfnamefont {H.~M.}\ \bibnamefont {{Mayer}}}, \bibinfo
  {author} {\bibfnamefont {U.}~\bibnamefont {{Ahlheim}}}, \ and\ \bibinfo
  {author} {\bibfnamefont {C.~D.}\ \bibnamefont {{Bredl}}},\ }\href@noop {}
  {\bibfield  {journal} {\bibinfo  {journal} {Z. Phys. B Cond. Matt.}\ }\textbf
  {\bibinfo {volume} {62}},\ \bibinfo {pages} {171} (\bibinfo {year}
  {1986})}\BibitemShut {NoStop}%
\bibitem [{\citenamefont {{Saran}}\ and\ \citenamefont
  {{McAlister}}(1988)}]{Saran88}%
  \BibitemOpen
  \bibfield  {author} {\bibinfo {author} {\bibfnamefont {M.}~\bibnamefont
  {{Saran}}}\ and\ \bibinfo {author} {\bibfnamefont {S.~P.}\ \bibnamefont
  {{McAlister}}},\ }\href@noop {} {\bibfield  {journal} {\bibinfo  {journal}
  {J. Magn. Magn. Mater.}\ }\textbf {\bibinfo {volume} {75}},\ \bibinfo {pages}
  {345} (\bibinfo {year} {1988})}\BibitemShut {NoStop}%
\bibitem [{\citenamefont {{Rebelsky}}\ \emph {et~al.}(1991)\citenamefont
  {{Rebelsky}}, \citenamefont {{McElfresh}}, \citenamefont {{Torikachvili}},\
  and\ \citenamefont {{Powell}}}]{Rebelsky91}%
  \BibitemOpen
  \bibfield  {author} {\bibinfo {author} {\bibfnamefont {L.}~\bibnamefont
  {{Rebelsky}}}, \bibinfo {author} {\bibfnamefont {M.~W.}\ \bibnamefont
  {{McElfresh}}}, \bibinfo {author} {\bibfnamefont {M.~S.}\ \bibnamefont
  {{Torikachvili}}}, \ and\ \bibinfo {author} {\bibfnamefont {B.~M.}\
  \bibnamefont {{Powell}}},\ }\href@noop {} {\bibfield  {journal} {\bibinfo
  {journal} {J. Appl. Phys.}\ }\textbf {\bibinfo {volume} {69}},\ \bibinfo
  {pages} {4810} (\bibinfo {year} {1991})}\BibitemShut {NoStop}%
\bibitem [{\citenamefont {{Torikachvili}}\ \emph {et~al.}(1992)\citenamefont
  {{Torikachvili}}, \citenamefont {{Jardim}}, \citenamefont {{Becerra}},
  \citenamefont {{Westphal}}, \citenamefont {{Paduan-Filho}}, \citenamefont
  {{Lopez}},\ and\ \citenamefont {{Rebelsky}}}]{Torikachvili92}%
  \BibitemOpen
  \bibfield  {author} {\bibinfo {author} {\bibfnamefont {M.~S.}\ \bibnamefont
  {{Torikachvili}}}, \bibinfo {author} {\bibfnamefont {R.~F.}\ \bibnamefont
  {{Jardim}}}, \bibinfo {author} {\bibfnamefont {C.~C.}\ \bibnamefont
  {{Becerra}}}, \bibinfo {author} {\bibfnamefont {C.~H.}\ \bibnamefont
  {{Westphal}}}, \bibinfo {author} {\bibfnamefont {A.}~\bibnamefont
  {{Paduan-Filho}}}, \bibinfo {author} {\bibfnamefont {V.~M.}\ \bibnamefont
  {{Lopez}}}, \ and\ \bibinfo {author} {\bibfnamefont {L.}~\bibnamefont
  {{Rebelsky}}},\ }\href@noop {} {\bibfield  {journal} {\bibinfo  {journal} {J.
  Magn. Magn. Mater.}\ }\textbf {\bibinfo {volume} {104-107}},\ \bibinfo
  {pages} {69} (\bibinfo {year} {1992})}\BibitemShut {NoStop}%
\bibitem [{\citenamefont {{Izumi}}\ and\ \citenamefont
  {{Momma}}(2007)}]{Izumi07}%
  \BibitemOpen
  \bibfield  {author} {\bibinfo {author} {\bibfnamefont {F.}~\bibnamefont
  {{Izumi}}}\ and\ \bibinfo {author} {\bibfnamefont {K.}~\bibnamefont
  {{Momma}}},\ }\href@noop {} {\bibfield  {journal} {\bibinfo  {journal} {Solid
  State Phenom.}\ }\textbf {\bibinfo {volume} {130}},\ \bibinfo {pages} {15}
  (\bibinfo {year} {2007})}\BibitemShut {NoStop}%
\bibitem [{\citenamefont {{Sechovsk\'{y}}}\ and\ \citenamefont
  {{Havela}}(1998)}]{Sechovsky98}%
  \BibitemOpen
  \bibfield  {author} {\bibinfo {author} {\bibfnamefont {V.}~\bibnamefont
  {{Sechovsk\'{y}}}}\ and\ \bibinfo {author} {\bibfnamefont {L.}~\bibnamefont
  {{Havela}}},\ }\href@noop {} {\emph {\bibinfo {title} {Handbook of Magnetic
  Materials}}},\ Vol.~\bibinfo {volume} {11}\ (\bibinfo  {publisher} {K. H. J.
  Bushcow (Ed.)},\ \bibinfo {year} {1998})\BibitemShut {NoStop}%
\bibitem [{\citenamefont {{McElfresh}}\ \emph {et~al.}(1987)\citenamefont
  {{McElfresh}}, \citenamefont {{Thompson}}, \citenamefont {{Willis}},
  \citenamefont {{Maple}}, \citenamefont {{Kohara}},\ and\ \citenamefont
  {{Torikachvili}}}]{McElfresh87}%
  \BibitemOpen
  \bibfield  {author} {\bibinfo {author} {\bibfnamefont {M.~W.}\ \bibnamefont
  {{McElfresh}}}, \bibinfo {author} {\bibfnamefont {J.~D.}\ \bibnamefont
  {{Thompson}}}, \bibinfo {author} {\bibfnamefont {J.~O.}\ \bibnamefont
  {{Willis}}}, \bibinfo {author} {\bibfnamefont {M.~B.}\ \bibnamefont
  {{Maple}}}, \bibinfo {author} {\bibfnamefont {T.}~\bibnamefont {{Kohara}}}, \
  and\ \bibinfo {author} {\bibfnamefont {M.~S.}\ \bibnamefont
  {{Torikachvili}}},\ }\href@noop {} {\bibfield  {journal} {\bibinfo  {journal}
  {Phys. Rev. B}\ }\textbf {\bibinfo {volume} {35}},\ \bibinfo {pages} {43}
  (\bibinfo {year} {1987})}\BibitemShut {NoStop}%
\bibitem [{\citenamefont {{Mihalik}}\ \emph {et~al.}(2007)\citenamefont
  {{Mihalik}}, \citenamefont {{Kolomiyets}}, \citenamefont {{Griveau}},
  \citenamefont {{Andreev}},\ and\ \citenamefont
  {{Sechovsk{\'y}}}}]{Mihalik07}%
  \BibitemOpen
  \bibfield  {author} {\bibinfo {author} {\bibfnamefont {M.}~\bibnamefont
  {{Mihalik}}}, \bibinfo {author} {\bibfnamefont {O.}~\bibnamefont
  {{Kolomiyets}}}, \bibinfo {author} {\bibfnamefont {J.-C.}\ \bibnamefont
  {{Griveau}}}, \bibinfo {author} {\bibfnamefont {A.~V.}\ \bibnamefont
  {{Andreev}}}, \ and\ \bibinfo {author} {\bibfnamefont {V.}~\bibnamefont
  {{Sechovsk{\'y}}}},\ }\href@noop {} {\bibfield  {journal} {\bibinfo
  {journal} {J. Phys. Soc. Jpn.}\ }\textbf {\bibinfo {volume} {76}},\ \bibinfo
  {pages} {54} (\bibinfo {year} {2007})}\BibitemShut {NoStop}%
\bibitem [{\citenamefont {{Ning}}\ \emph {et~al.}(1992)\citenamefont {{Ning}},
  \citenamefont {{Garrett}}, \citenamefont {{Stager}},\ and\ \citenamefont
  {{Datars}}}]{Ning92}%
  \BibitemOpen
  \bibfield  {author} {\bibinfo {author} {\bibfnamefont {Y.~B.}\ \bibnamefont
  {{Ning}}}, \bibinfo {author} {\bibfnamefont {J.~D.}\ \bibnamefont
  {{Garrett}}}, \bibinfo {author} {\bibfnamefont {C.~V.}\ \bibnamefont
  {{Stager}}}, \ and\ \bibinfo {author} {\bibfnamefont {W.~R.}\ \bibnamefont
  {{Datars}}},\ }\href@noop {} {\bibfield  {journal} {\bibinfo  {journal}
  {Phys. Rev. B}\ }\textbf {\bibinfo {volume} {46}},\ \bibinfo {pages} {8201}
  (\bibinfo {year} {1992})}\BibitemShut {NoStop}%
\bibitem [{\citenamefont {Anderson}\ and\ \citenamefont
  {Callen}(1964)}]{Burr64}%
  \BibitemOpen
  \bibfield  {author} {\bibinfo {author} {\bibfnamefont {F.~B.}\ \bibnamefont
  {Anderson}}\ and\ \bibinfo {author} {\bibfnamefont {H.~B.}\ \bibnamefont
  {Callen}},\ }\href {\doibase 10.1103/PhysRev.136.A1068} {\bibfield  {journal}
  {\bibinfo  {journal} {Phys. Rev.}\ }\textbf {\bibinfo {volume} {136}},\
  \bibinfo {pages} {A1068} (\bibinfo {year} {1964})}\BibitemShut {NoStop}%
\bibitem [{\citenamefont {{Schmidt}}\ and\ \citenamefont
  {{Friedberg}}(1967)}]{Schmidt67}%
  \BibitemOpen
  \bibfield  {author} {\bibinfo {author} {\bibfnamefont {V.~A.}\ \bibnamefont
  {{Schmidt}}}\ and\ \bibinfo {author} {\bibfnamefont {S.~A.}\ \bibnamefont
  {{Friedberg}}},\ }\href {\doibase http://dx.doi.org/10.1063/1.1709322}
  {\bibfield  {journal} {\bibinfo  {journal} {Journal of Applied Physics}\
  }\textbf {\bibinfo {volume} {38}},\ \bibinfo {pages} {5319} (\bibinfo {year}
  {1967})}\BibitemShut {NoStop}%
\bibitem [{\citenamefont {Schulze~Grachtrup}\ \emph {et~al.}(2012)\citenamefont
  {Schulze~Grachtrup}, \citenamefont {Bleckmann}, \citenamefont {Willenberg},
  \citenamefont {S\"ullow}, \citenamefont {Bartkowiak}, \citenamefont
  {Skourski}, \citenamefont {Rakoto}, \citenamefont {Sheikin},\ and\
  \citenamefont {Mydosh}}]{Schulze12}%
  \BibitemOpen
  \bibfield  {author} {\bibinfo {author} {\bibfnamefont {D.}~\bibnamefont
  {Schulze~Grachtrup}}, \bibinfo {author} {\bibfnamefont {M.}~\bibnamefont
  {Bleckmann}}, \bibinfo {author} {\bibfnamefont {B.}~\bibnamefont
  {Willenberg}}, \bibinfo {author} {\bibfnamefont {S.}~\bibnamefont
  {S\"ullow}}, \bibinfo {author} {\bibfnamefont {M.}~\bibnamefont
  {Bartkowiak}}, \bibinfo {author} {\bibfnamefont {Y.}~\bibnamefont
  {Skourski}}, \bibinfo {author} {\bibfnamefont {H.}~\bibnamefont {Rakoto}},
  \bibinfo {author} {\bibfnamefont {I.}~\bibnamefont {Sheikin}}, \ and\
  \bibinfo {author} {\bibfnamefont {J.~A.}\ \bibnamefont {Mydosh}},\ }\href
  {\doibase 10.1103/PhysRevB.85.054410} {\bibfield  {journal} {\bibinfo
  {journal} {Phys. Rev. B}\ }\textbf {\bibinfo {volume} {85}},\ \bibinfo
  {pages} {054410} (\bibinfo {year} {2012})}\BibitemShut {NoStop}%
\bibitem [{\citenamefont {Sakurazawa}\ \emph {et~al.}(2005)\citenamefont
  {Sakurazawa}, \citenamefont {Kontani},\ and\ \citenamefont
  {Saso}}]{Sakurazawa05}%
  \BibitemOpen
  \bibfield  {author} {\bibinfo {author} {\bibfnamefont {K.}~\bibnamefont
  {Sakurazawa}}, \bibinfo {author} {\bibfnamefont {H.}~\bibnamefont {Kontani}},
  \ and\ \bibinfo {author} {\bibfnamefont {T.}~\bibnamefont {Saso}},\ }\href
  {\doibase 10.1143/JPSJ.74.271} {\bibfield  {journal} {\bibinfo  {journal}
  {Journal of the Physical Society of Japan}\ }\textbf {\bibinfo {volume}
  {74}},\ \bibinfo {pages} {271} (\bibinfo {year} {2005})},\ \Eprint
  {http://arxiv.org/abs/http://dx.doi.org/10.1143/JPSJ.74.271}
  {http://dx.doi.org/10.1143/JPSJ.74.271} \BibitemShut {NoStop}%
\bibitem [{\citenamefont {Knebel}\ \emph {et~al.}(2011)\citenamefont {Knebel},
  \citenamefont {Buhot}, \citenamefont {Aoki}, \citenamefont {Lapertot},
  \citenamefont {Raymond}, \citenamefont {Ressouche},\ and\ \citenamefont
  {Flouquet}}]{Knebel11}%
  \BibitemOpen
  \bibfield  {author} {\bibinfo {author} {\bibfnamefont {G.}~\bibnamefont
  {Knebel}}, \bibinfo {author} {\bibfnamefont {J.}~\bibnamefont {Buhot}},
  \bibinfo {author} {\bibfnamefont {D.}~\bibnamefont {Aoki}}, \bibinfo {author}
  {\bibfnamefont {G.}~\bibnamefont {Lapertot}}, \bibinfo {author}
  {\bibfnamefont {S.}~\bibnamefont {Raymond}}, \bibinfo {author} {\bibfnamefont
  {E.}~\bibnamefont {Ressouche}}, \ and\ \bibinfo {author} {\bibfnamefont
  {J.}~\bibnamefont {Flouquet}},\ }\href {\doibase 10.1143/JPSJS.80SA.SA001}
  {\bibfield  {journal} {\bibinfo  {journal} {Journal of the Physical Society
  of Japan}\ }\textbf {\bibinfo {volume} {80}},\ \bibinfo {pages} {SA001}
  (\bibinfo {year} {2011})},\ \Eprint
  {http://arxiv.org/abs/http://dx.doi.org/10.1143/JPSJS.80SA.SA001}
  {http://dx.doi.org/10.1143/JPSJS.80SA.SA001} \BibitemShut {NoStop}%
\bibitem [{\citenamefont {{Schmidt, Burkhard}}\ \emph
  {et~al.}(2013)\citenamefont {{Schmidt, Burkhard}}, \citenamefont {{Siahatgar,
  Mohammad}},\ and\ \citenamefont {{Thalmeier, Peter}}}]{Schmidt13}%
  \BibitemOpen
  \bibfield  {author} {\bibinfo {author} {\bibnamefont {{Schmidt, Burkhard}}},
  \bibinfo {author} {\bibnamefont {{Siahatgar, Mohammad}}}, \ and\ \bibinfo
  {author} {\bibnamefont {{Thalmeier, Peter}}},\ }\href {\doibase
  10.1051/epjconf/20134004001} {\bibfield  {journal} {\bibinfo  {journal} {EPJ
  Web of Conferences}\ }\textbf {\bibinfo {volume} {40}},\ \bibinfo {pages}
  {04001} (\bibinfo {year} {2013})}\BibitemShut {NoStop}%
\bibitem [{\citenamefont {Fortune}\ \emph {et~al.}(2014)\citenamefont
  {Fortune}, \citenamefont {Hannahs}, \citenamefont {Landee}, \citenamefont
  {Turnbull},\ and\ \citenamefont {Xiao}}]{Fortune14}%
  \BibitemOpen
  \bibfield  {author} {\bibinfo {author} {\bibfnamefont {N.~A.}\ \bibnamefont
  {Fortune}}, \bibinfo {author} {\bibfnamefont {S.~T.}\ \bibnamefont
  {Hannahs}}, \bibinfo {author} {\bibfnamefont {C.~P.}\ \bibnamefont {Landee}},
  \bibinfo {author} {\bibfnamefont {M.~M.}\ \bibnamefont {Turnbull}}, \ and\
  \bibinfo {author} {\bibfnamefont {F.}~\bibnamefont {Xiao}},\ }\href
  {http://stacks.iop.org/1742-6596/568/i=4/a=042004} {\bibfield  {journal}
  {\bibinfo  {journal} {Journal of Physics: Conference Series}\ }\textbf
  {\bibinfo {volume} {568}},\ \bibinfo {pages} {042004} (\bibinfo {year}
  {2014})}\BibitemShut {NoStop}%
\bibitem [{\citenamefont {{Takacs}}\ and\ \citenamefont {{M{\' e}sz{\'
  a}ros}}(2008)}]{Takacs08}%
  \BibitemOpen
  \bibfield  {author} {\bibinfo {author} {\bibfnamefont {J.}~\bibnamefont
  {{Takacs}}}\ and\ \bibinfo {author} {\bibfnamefont {I.}~\bibnamefont {{M{\'
  e}sz{\' a}ros}}},\ }\href@noop {} {\bibfield  {journal} {\bibinfo  {journal}
  {Physica B}\ }\textbf {\bibinfo {volume} {403}},\ \bibinfo {pages} {3137}
  (\bibinfo {year} {2008})}\BibitemShut {NoStop}%
\bibitem [{\citenamefont {{K{\"u}tterer}}\ \emph {et~al.}(1977)\citenamefont
  {{K{\"u}tterer}}, \citenamefont {{Hilzinger}},\ and\ \citenamefont
  {{Kronm{\"u}ller}}}]{Kutterer77}%
  \BibitemOpen
  \bibfield  {author} {\bibinfo {author} {\bibfnamefont {R.}~\bibnamefont
  {{K{\"u}tterer}}}, \bibinfo {author} {\bibfnamefont {H.-R.}\ \bibnamefont
  {{Hilzinger}}}, \ and\ \bibinfo {author} {\bibfnamefont {H.}~\bibnamefont
  {{Kronm{\"u}ller}}},\ }\href@noop {} {\bibfield  {journal} {\bibinfo
  {journal} {J. Magn. Magn. Matter.}\ }\textbf {\bibinfo {volume} {4}},\
  \bibinfo {pages} {1} (\bibinfo {year} {1977})}\BibitemShut {NoStop}%
\bibitem [{\citenamefont {{Ishii}}\ \emph {et~al.}(2008)\citenamefont
  {{Ishii}}, \citenamefont {{Okamura}}, \citenamefont {{Chiba}}, \citenamefont
  {{Nogami}},\ and\ \citenamefont {{Ishida}}}]{Ishii08}%
  \BibitemOpen
  \bibfield  {author} {\bibinfo {author} {\bibfnamefont {N.}~\bibnamefont
  {{Ishii}}}, \bibinfo {author} {\bibfnamefont {Y.}~\bibnamefont {{Okamura}}},
  \bibinfo {author} {\bibfnamefont {S.}~\bibnamefont {{Chiba}}}, \bibinfo
  {author} {\bibfnamefont {T.}~\bibnamefont {{Nogami}}}, \ and\ \bibinfo
  {author} {\bibfnamefont {T.}~\bibnamefont {{Ishida}}},\ }\href@noop {}
  {\bibfield  {journal} {\bibinfo  {journal} {J. Am. Chem. Soc.}\ }\textbf
  {\bibinfo {volume} {130}},\ \bibinfo {pages} {24} (\bibinfo {year}
  {2008})}\BibitemShut {NoStop}%
\bibitem [{\citenamefont {{Yokoyama}}\ \emph {et~al.}(2005)\citenamefont
  {{Yokoyama}}, \citenamefont {{Amitsuka}}, \citenamefont {{Tenya}},
  \citenamefont {{Watanabe}}, \citenamefont {{Kawarazaki}}, \citenamefont
  {{Yoshizawa}},\ and\ \citenamefont {{Mydosh}}}]{Yokoyama05}%
  \BibitemOpen
  \bibfield  {author} {\bibinfo {author} {\bibfnamefont {M.}~\bibnamefont
  {{Yokoyama}}}, \bibinfo {author} {\bibfnamefont {H.}~\bibnamefont
  {{Amitsuka}}}, \bibinfo {author} {\bibfnamefont {K.}~\bibnamefont {{Tenya}}},
  \bibinfo {author} {\bibfnamefont {K.}~\bibnamefont {{Watanabe}}}, \bibinfo
  {author} {\bibfnamefont {S.}~\bibnamefont {{Kawarazaki}}}, \bibinfo {author}
  {\bibfnamefont {H.}~\bibnamefont {{Yoshizawa}}}, \ and\ \bibinfo {author}
  {\bibfnamefont {J.~A.}\ \bibnamefont {{Mydosh}}},\ }\href@noop {} {\bibfield
  {journal} {\bibinfo  {journal} {Phys. Rev. B}\ }\textbf {\bibinfo {volume}
  {72}},\ \bibinfo {pages} {214419} (\bibinfo {year} {2005})}\BibitemShut
  {NoStop}%
\bibitem [{\citenamefont {{Amitsuka}}\ \emph {et~al.}(2007)\citenamefont
  {{Amitsuka}}, \citenamefont {{Matsuda}}, \citenamefont {{Kawasaki}},
  \citenamefont {{Yokoyama}}, \citenamefont {{Sekine}}, \citenamefont
  {{Tateiwa}}, \citenamefont {{Kobayashi}}, \citenamefont {{Kawarazaki}},\ and\
  \citenamefont {{Yoshizawa}}}]{Amitsuka07}%
  \BibitemOpen
  \bibfield  {author} {\bibinfo {author} {\bibfnamefont {H.}~\bibnamefont
  {{Amitsuka}}}, \bibinfo {author} {\bibfnamefont {K.}~\bibnamefont
  {{Matsuda}}}, \bibinfo {author} {\bibfnamefont {K.}~\bibnamefont
  {{Kawasaki}}, \bibfnamefont {I.~andd~{Tenya}}}, \bibinfo {author}
  {\bibfnamefont {M.}~\bibnamefont {{Yokoyama}}}, \bibinfo {author}
  {\bibfnamefont {C.}~\bibnamefont {{Sekine}}}, \bibinfo {author}
  {\bibfnamefont {N.}~\bibnamefont {{Tateiwa}}}, \bibinfo {author}
  {\bibfnamefont {T.~C.}\ \bibnamefont {{Kobayashi}}}, \bibinfo {author}
  {\bibfnamefont {S.}~\bibnamefont {{Kawarazaki}}}, \ and\ \bibinfo {author}
  {\bibfnamefont {H.}~\bibnamefont {{Yoshizawa}}},\ }\href@noop {} {\bibfield
  {journal} {\bibinfo  {journal} {J. Magn. Magn. Mater.}\ }\textbf {\bibinfo
  {volume} {310}},\ \bibinfo {pages} {214} (\bibinfo {year}
  {2007})}\BibitemShut {NoStop}%
\bibitem [{\citenamefont {{Kuwahara}}\ \emph {et~al.}(2013)\citenamefont
  {{Kuwahara}}, \citenamefont {{Yoshii}}, \citenamefont {{Nojiri}},
  \citenamefont {{Aoki}}, \citenamefont {{Knafo}}, \citenamefont {{Duc}},
  \citenamefont {{Fabreges}}, \citenamefont {{Scheerer}}, \citenamefont
  {{Frings}}, \citenamefont {{Rikken}}, \citenamefont {{Bourdarot}},
  \citenamefont {{Regnault}},\ and\ \citenamefont {{Flouquet}}}]{Kuwahara13}%
  \BibitemOpen
  \bibfield  {author} {\bibinfo {author} {\bibfnamefont {K.}~\bibnamefont
  {{Kuwahara}}}, \bibinfo {author} {\bibfnamefont {S.}~\bibnamefont
  {{Yoshii}}}, \bibinfo {author} {\bibfnamefont {H.}~\bibnamefont {{Nojiri}}},
  \bibinfo {author} {\bibfnamefont {D.}~\bibnamefont {{Aoki}}}, \bibinfo
  {author} {\bibfnamefont {W.}~\bibnamefont {{Knafo}}}, \bibinfo {author}
  {\bibfnamefont {F.}~\bibnamefont {{Duc}}}, \bibinfo {author} {\bibfnamefont
  {X.}~\bibnamefont {{Fabreges}}}, \bibinfo {author} {\bibfnamefont {G.~W.}\
  \bibnamefont {{Scheerer}}}, \bibinfo {author} {\bibfnamefont
  {P.}~\bibnamefont {{Frings}}}, \bibinfo {author} {\bibfnamefont {G.~L.
  J.~A.}\ \bibnamefont {{Rikken}}}, \bibinfo {author} {\bibfnamefont
  {F.}~\bibnamefont {{Bourdarot}}}, \bibinfo {author} {\bibfnamefont {L.~P.}\
  \bibnamefont {{Regnault}}}, \ and\ \bibinfo {author} {\bibfnamefont
  {J.}~\bibnamefont {{Flouquet}}},\ }\href@noop {} {\bibfield  {journal}
  {\bibinfo  {journal} {Phys. Rev. Lett.}\ }\textbf {\bibinfo {volume} {110}},\
  \bibinfo {pages} {216406} (\bibinfo {year} {2013})}\BibitemShut {NoStop}%
\bibitem [{\citenamefont {{Honma}}\ \emph {et~al.}(2015)\citenamefont
  {{Honma}}, \citenamefont {{Amitsuka}}, \citenamefont {{Sakakibara}},
  \citenamefont {{Sugiyama}},\ and\ \citenamefont {{Date}}}]{Honma93}%
  \BibitemOpen
  \bibfield  {author} {\bibinfo {author} {\bibfnamefont {T.}~\bibnamefont
  {{Honma}}}, \bibinfo {author} {\bibfnamefont {H.}~\bibnamefont {{Amitsuka}}},
  \bibinfo {author} {\bibfnamefont {T.}~\bibnamefont {{Sakakibara}}}, \bibinfo
  {author} {\bibfnamefont {K.}~\bibnamefont {{Sugiyama}}}, \ and\ \bibinfo
  {author} {\bibfnamefont {M.}~\bibnamefont {{Date}}},\ }\href@noop {}
  {\bibfield  {journal} {\bibinfo  {journal} {Physica B}\ }\textbf {\bibinfo
  {volume} {186-188}},\ \bibinfo {pages} {684} (\bibinfo {year}
  {2015})}\BibitemShut {NoStop}%
\bibitem [{\citenamefont {{Herbst}}(1991)}]{Herbst91}%
  \BibitemOpen
  \bibfield  {author} {\bibinfo {author} {\bibfnamefont {J.~F.}\ \bibnamefont
  {{Herbst}}},\ }\href@noop {} {\bibfield  {journal} {\bibinfo  {journal} {Rev.
  Mod. Phys.}\ }\textbf {\bibinfo {volume} {63}},\ \bibinfo {pages} {819}
  (\bibinfo {year} {1991})}\BibitemShut {NoStop}%
\bibitem [{\citenamefont {{Koelling}}\ \emph {et~al.}(1985)\citenamefont
  {{Koelling}}, \citenamefont {{Dunlap}},\ and\ \citenamefont
  {{Crabtree}}}]{Koelling85}%
  \BibitemOpen
  \bibfield  {author} {\bibinfo {author} {\bibfnamefont {D.~D.}\ \bibnamefont
  {{Koelling}}}, \bibinfo {author} {\bibfnamefont {B.~D.}\ \bibnamefont
  {{Dunlap}}}, \ and\ \bibinfo {author} {\bibfnamefont {G.~W.}\ \bibnamefont
  {{Crabtree}}},\ }\href@noop {} {\bibfield  {journal} {\bibinfo  {journal}
  {Phys. Rev. B}\ }\textbf {\bibinfo {volume} {31}},\ \bibinfo {pages} {4966}
  (\bibinfo {year} {1985})}\BibitemShut {NoStop}%
\bibitem [{\citenamefont {{Cooper}}\ \emph {et~al.}(1987)\citenamefont
  {{Cooper}}, \citenamefont {{Hu}}, \citenamefont {{Kioussis}},\ and\
  \citenamefont {{Wills}}}]{Cooper87}%
  \BibitemOpen
  \bibfield  {author} {\bibinfo {author} {\bibfnamefont {B.~R.}\ \bibnamefont
  {{Cooper}}}, \bibinfo {author} {\bibfnamefont {G.-J.}\ \bibnamefont {{Hu}}},
  \bibinfo {author} {\bibfnamefont {N.}~\bibnamefont {{Kioussis}}}, \ and\
  \bibinfo {author} {\bibfnamefont {J.~M.}\ \bibnamefont {{Wills}}},\
  }\href@noop {} {\bibfield  {journal} {\bibinfo  {journal} {J. Magn. Magn.
  Mater.}\ }\textbf {\bibinfo {volume} {63-64}},\ \bibinfo {pages} {121}
  (\bibinfo {year} {1987})}\BibitemShut {NoStop}%
\bibitem [{\citenamefont {{Coqblin}}\ and\ \citenamefont
  {{Schrieffer}}(1969)}]{Coqblin69}%
  \BibitemOpen
  \bibfield  {author} {\bibinfo {author} {\bibfnamefont {B.}~\bibnamefont
  {{Coqblin}}}\ and\ \bibinfo {author} {\bibfnamefont {J.~R.}\ \bibnamefont
  {{Schrieffer}}},\ }\href@noop {} {\bibfield  {journal} {\bibinfo  {journal}
  {Phys. Rev.}\ }\textbf {\bibinfo {volume} {185}},\ \bibinfo {pages} {847}
  (\bibinfo {year} {1969})}\BibitemShut {NoStop}%
\bibitem [{\citenamefont {{Hu}}\ and\ \citenamefont {{Cooper}}(1993)}]{Hu93}%
  \BibitemOpen
  \bibfield  {author} {\bibinfo {author} {\bibfnamefont {G.-J.}\ \bibnamefont
  {{Hu}}}\ and\ \bibinfo {author} {\bibfnamefont {B.~R.}\ \bibnamefont
  {{Cooper}}},\ }\href@noop {} {\bibfield  {journal} {\bibinfo  {journal}
  {Phys. Rev. B}\ }\textbf {\bibinfo {volume} {48}},\ \bibinfo {pages} {12743}
  (\bibinfo {year} {1993})}\BibitemShut {NoStop}%
\bibitem [{\citenamefont {{Endstra}}\ \emph {et~al.}(1993)\citenamefont
  {{Endstra}}, \citenamefont {{Nieuwenhuys}},\ and\ \citenamefont
  {{Mydosh}}}]{Endstra93}%
  \BibitemOpen
  \bibfield  {author} {\bibinfo {author} {\bibfnamefont {T.}~\bibnamefont
  {{Endstra}}}, \bibinfo {author} {\bibfnamefont {G.~J.}\ \bibnamefont
  {{Nieuwenhuys}}}, \ and\ \bibinfo {author} {\bibfnamefont {J.~A.}\
  \bibnamefont {{Mydosh}}},\ }\href@noop {} {\bibfield  {journal} {\bibinfo
  {journal} {Phys. Rev. B}\ }\textbf {\bibinfo {volume} {48}},\ \bibinfo
  {pages} {9595} (\bibinfo {year} {1993})}\BibitemShut {NoStop}%
\bibitem [{\citenamefont {{Sandratskii}}\ and\ \citenamefont
  {{Kubler}}(1994)}]{Sandratskii94}%
  \BibitemOpen
  \bibfield  {author} {\bibinfo {author} {\bibfnamefont {L.~M.}\ \bibnamefont
  {{Sandratskii}}}\ and\ \bibinfo {author} {\bibfnamefont {J.}~\bibnamefont
  {{Kubler}}},\ }\href@noop {} {\bibfield  {journal} {\bibinfo  {journal}
  {Phys. Rev. B}\ }\textbf {\bibinfo {volume} {50}},\ \bibinfo {pages} {9258}
  (\bibinfo {year} {1994})}\BibitemShut {NoStop}%
\end{thebibliography}
\end{document}